\documentclass[%
,aps%
 ,twocolumn%
 ,secnumarabic%
,amssymb, amsmath,nobibnotes, aps, prd, floatfix]{revtex4}
\usepackage[utf8]{inputenc}
\usepackage{ marvosym }
\usepackage{graphicx}
\usepackage[compat=1.1.0]{tikz-feynman}
\usepackage{subcaption}
\usepackage{hyperref}
\usepackage{comment}
\usepackage{float}
\usepackage{scalerel}
\usepackage{slashed}

\usepackage{tikz}
\usetikzlibrary{decorations.pathmorphing}
\usetikzlibrary{decorations.markings}
\tikzset{zigzag/.style={decorate, decoration=zigzag}}

\usepackage{latexsym,graphicx,color,amsmath}
\usepackage{amssymb,mathrsfs}

\def\<{\langle} 			
\def\>{\rangle}

\newcommand\bcc{\begin{pmatrix}}
\newcommand\ecc{\end{pmatrix}}

\DeclareMathOperator*{\argmin}{arg\,min}

\begin{document}


\title{Automated label flows for excited states of correlation functions in lattice gauge theory}


\author{Kimmy K. Cushman}
\email[]{kimmy.cushman@yale.edu}
\affiliation{Department of Physics, Yale University, 217 Prospect Street, New Haven, Connecticut 06511, USA}

\author{George T. Fleming}
\email[]{george.fleming@yale.edu}
\affiliation{Department of Physics, Yale University, 217 Prospect Street, New Haven, Connecticut 06511, USA}


\date{\today}

\begin{abstract}
Extracting excited states from lattice gauge theory correlation functions can be achieved through chi-squared minimization fits or algebraic approaches such as the variational method and Prony's method. Performing any kind of error analysis, such as bootstrap resampling, often leads to overlapping confidence regions of model parameters, even when the spectrum is not particularly dense. In order to correctly estimate errors, one must beware of mislabeling the states. In this work, we provide an algorithm that we call {\it automated label flows} which consistently and systematically identifies a deterministic labeling of states. In the context of Prony's method, we analyze lattice correlation functions by using automated label flows, and compare the results to fits obtained from chi-square minimization fits to exponentials.
\end{abstract}

\keywords{lattice field theory, lattice QCD, black-box method, correlation functions, bootstrapping, Prony's method}

\maketitle

\section{Introduction}
Correlation functions in lattice field theory encode information about the eigenvalues and eigenvectors of the Hamiltonian of a theory. Often only the ground state energies and matrix elements are of interest, but a tower of excited states contaminates the signal in any realistic lattice calculation. Modeling the first excited state well helps produce a better variational estimate of the ground state, so extracting the first few excited states is a common practice. For example, lattice spectroscopy of nuclei requires a robust method for extracting excited states because their spectrum is dense compared to hadrons~\cite{nuclear1, nuc_spec}.  Additionally, the excited state spectrum may be of interest because it encodes two-particle scattering phase shifts through the L{\"u}scher method~\cite{Davoudi:2018wgb,Luscher1}. Similarly, extracting excited states from confining gauge theories is crucial for beyond the Standard Model lattice spectroscopy in explorations such as composite Higgs~\cite{Oliver, Appelquist:2018yqe}.

Various methods are used for the extraction of excited states from lattice correlation functions, many of which involve utilizing multiple correlation functions from a given quantum channel. These can be classified into two types: least squares approaches and algebraic approaches. Fitting to a sum of exponentials is a common practice for single or multiple correlation functions using a joint chi-square minimization (see for example~\cite{Degrand_book}). When correlation matrices are available, the variational method is a common algebraic approach to spectroscopy~\cite{Berg:1982kp}. Additionally, Matrix Prony has been used in a similar vein~\cite{Berkowitz:2019yrf, Beane:2010em}. Also, recent studies~\cite{2019arXiv190210695C, 2019arXiv190701547K} have discussed an algebraic approach known as Prony's method~\cite{Prony,phdthesis}, where $M$ energies are obtained by finding the roots of an $M^{\rm th}$ order polynomial constructed from a Hankel matrix of the data~\cite{Fleming:2009wb, Fleming:2004hs, Fleming2010}.

In any method, the best-fit values of the masses and matrix elements are only single points in parameter space, but to determine errors on the parameters, a confidence region must be established for each point. Therefore, regardless of the method, error estimation can result in states that have overlapping confidence regions, making identification of states ambiguous. In most methods, this may be a result of bootstrap or jackknife resampling, whereas in least squares fitting, it may also appear when modeling contours of $\chi^2$ fits to parameters.  Error estimation is especially problematic at large Euclidean times where the signal to noise may be low, as well as when many states are extracted. By using multiple correlation functions, this problem can be mediated with great effort and expense by operator pruning, where the operator basis of a correlation function matrix is carefully chosen. Such an approach is expensive in both the computation of the correlation functions and the user input required for pruning. In this work, instead of avoiding the multi-state labeling problem, we attack it by introducing a method we call {\it automated label flows}. In Section 2, we outline Prony's method as the algebraic method we use, and depict the labeling problem as it emerges for real correlation function data. In Section 3, we describe the process of automated label flows, and show its use for our labeling problem. In Section 4 we compare the results of the multi-state extraction using automated label flows to those obtained from a chi-square minimization exponential fit, then we conclude in Section 5.

\section{Prony's Method}
In this section we outline the algebraic method for extracting $M$ states from correlation function data. The method can be generalized to a correlation matrix using Matrix Prony, but here we focus on extraction from a single correlation function in order to address the challenges of overlapping errors. 
\subsection{Prony's method in theory}
Consider a two point function (one element of a correlation matrix) of the following form : 
\begin{align}
C_{ij}(t) &= \<0|\mathcal{O}_i^{\dagger}(t) \mathcal{O}_j(0) |0\>\\
&= \sum_{m=1}^{\infty} \<0|\mathcal{O}_i^{\dagger}|E_m\> {\rm e}^{-E_m t}\<E_m|\mathcal{O}_j |0\> \\
\Rightarrow C_{ij}(t) &= \sum_{m=1}^{\infty} a_{ij,m} \,{\rm e}^{-E_m t}\label{corr}.
\end{align}
Note that the sum is invariant under permutation of labels of states, $m$. As we will see, this is problematic when the error regions of pairs $(a_m, E_m)$ overlap because the labeling can be ambiguous. Nevertheless, to extract M states, we model the correlation function as a sum of $M$ exponentials. An exactly constrained {\it stencil} of $2M$ timeslices from the correlation function can be written as a length $2M$ vector $y_{n}(t) \equiv C(t+n)$:
\begin{align}
y_n(t) & =  \sum_{m=1}^{M} a_m {\rm e}^{-E_m t} {\rm e}^{-E_m n} \\
&= \sum_{m=1}^{M} A_m(t) z_m^n,\label{linear}
\end{align}
where we define $A_m(t) \equiv a_m {\rm e}^{-E_m t}$ and $z_m \equiv {\rm e}^{-E_m}$. For each choice of stencil (each timeslices $t$), this equation can be written as the following matrix equation:
\begin{align}
\bcc y_0 \\ y_1 \\ \vdots \\ y_{\scaleto{2M-1}{4.5pt}} \ecc & = \bcc 1 & 		1  & \cdots & 1\\[0.1cm]
									z_1 &  z_2 & \cdots & z_{M}\\[0.1cm]
									z_1^2 &  z_2^2 & \cdots & z_M^2\\[0.1cm]
									\vdots & \vdots & \ddots & \vdots\\[0.1cm]
									z_1^{\scaleto{2M-1}{4.5pt}} &  z_2^{\scaleto{2M-1}{4.5pt}} & \cdots & z_{M}^{\scaleto{2M-1}{4.5pt}}\\[0.1cm]
							\ecc  \bcc A_1 \\ A_2 \\ \vdots \\ A_{M} \ecc .\label{vandermonde}
\end{align}
Prony's algebraic result~\cite{Prony, phdthesis} to solve a system of this form can be used to solve for the $z$'s algebraically by finding the roots of the $M^{\rm th}$ order polynomial from the so-called Hankel matrix determinant,
\begin{align}
0 & = \begin{vmatrix}
y_0 		& y_1 	& \cdots &y_{M-2} 	& y_{M-1} 		& 1\\
y_1 		& y_2 	& \cdots & y_{M-1} 	& y_{M} 		& z\\
y_2 		& y_3 	& \cdots &y_M 		& y_{M+1}		& z^2\\
\vdots 	& \vdots 	& \ddots & \vdots 	&\vdots 		& \vdots \\
y_M 		& y_{M+1}& \cdots &y_{2M-2} 	& y_{2M-1}	& z^M
\end{vmatrix} \, \,.
\end{align}\\[0.2cm]
 Once the energies have been determined in this way, the amplitudes can be found by solving the linear system of Eq.~\ref{vandermonde}.

\subsection{Prony's method in practice with real data}
Using Prony's method as described above, we analyze a subset of correlation function data obtained from the LSD collaboration's investigation of an $SU(3)$ gauge theory with $N_f = 8$ degenerate flavors~\cite{Appelquist:2018yqe}. Throughout this work we show multiple state extractions for the pseudoscalar  point-point correlator. To perform an error estimation we use bootstrap resampling (see, for example~\cite{Effron_book}) of the mean of the 232 correlators, $C(t)$, with $t= 0,1,2,\ldots, N_t = 128$, computed in this dataset.

\begin{figure*}
\begin{subfigure}[t]{0.49\linewidth}
\includegraphics[width=1\linewidth]{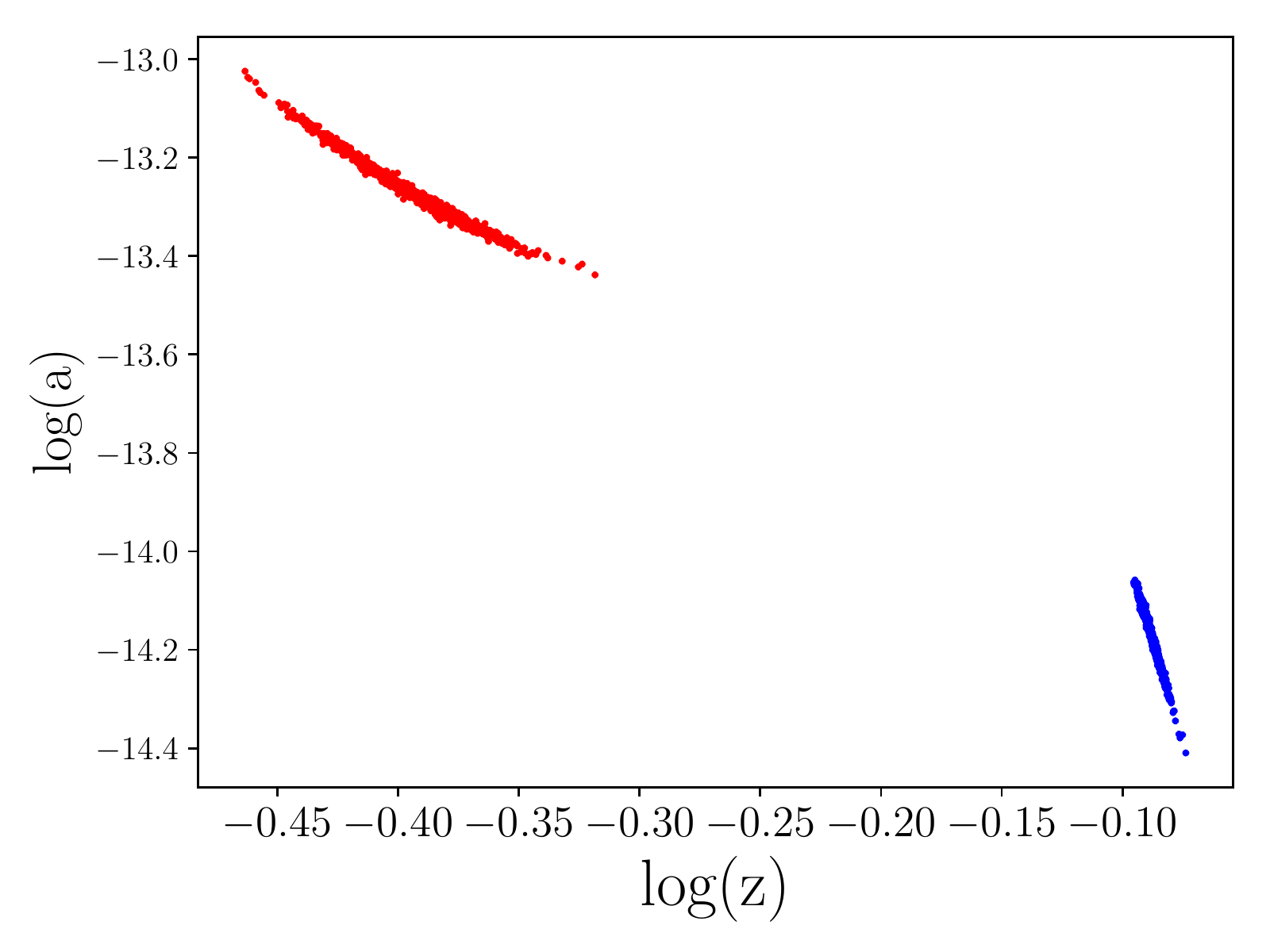}
\caption{Sets of $M=2$ tuples $(z_m, a_m)$ labeled according to their mass for 1000 bootstrap samples of $y_n(t=8)$ from Prony's method.}
\label{early_M2}
\end{subfigure}%
\hfill
\begin{subfigure}[t]{0.49\linewidth}
\includegraphics[width=1\linewidth]{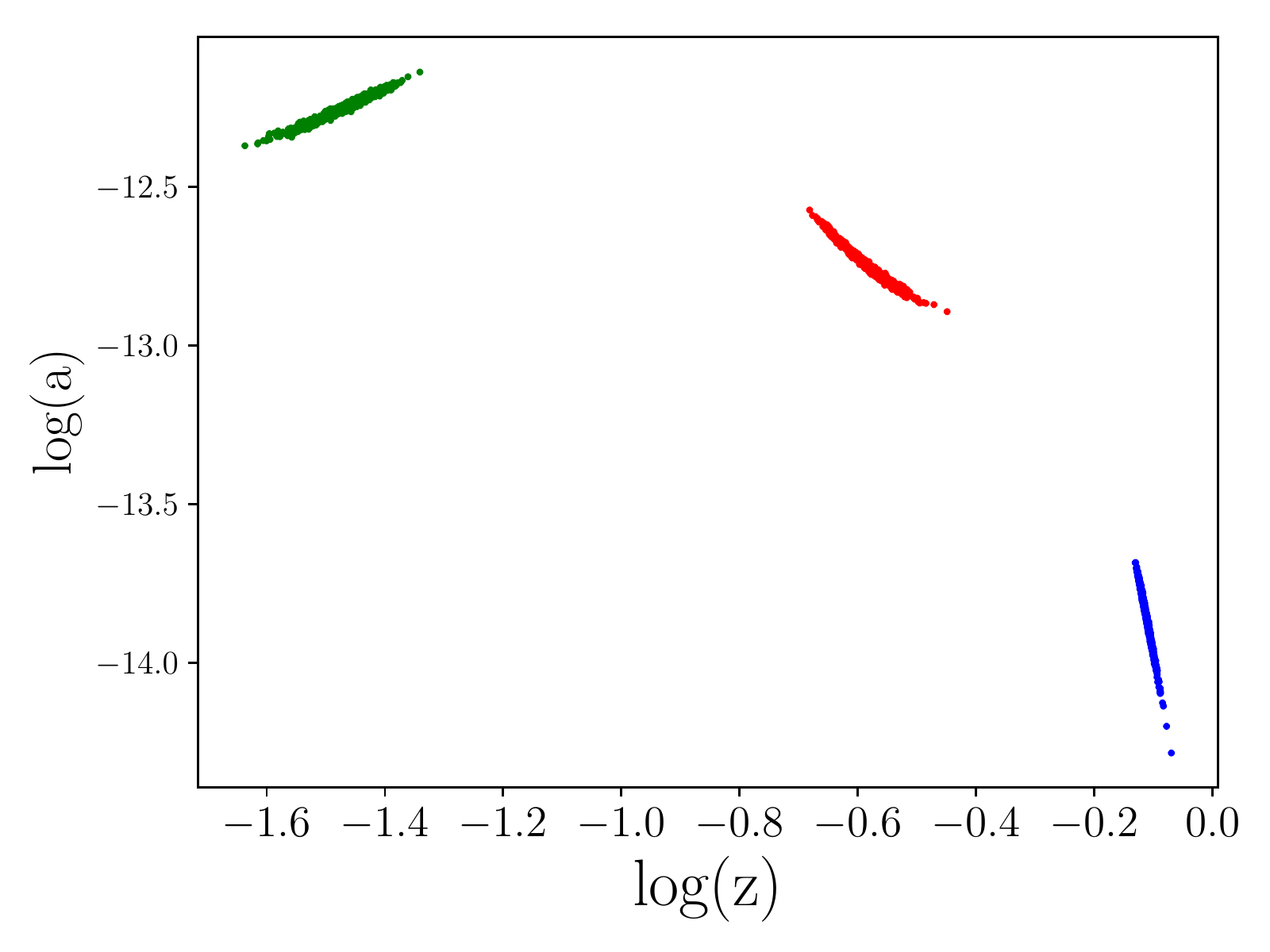}
\caption{Sets of $M=3$ tuples $(z_m, a_m)$ labeled according to their mass for 1000 bootstrap samples of $y_n(t=1)$ from Prony's method.}
\label{early_M3}
\end{subfigure}%
\caption{Labeling according to mass when data is noisy. Plot shows real part of logs.}
\end{figure*}

The problem of clustering emerges as the question of how to assign labels (ground state, first excited state, etc.) to each of the 1000 bootstrap sets of two ($M=2$) or three ($M=3$), etc. tuples $(z_m, a_m)$ extracted from Prony's method. At early times when there is little noise in the stencil of data $y_n(t)$, there is also a small variance in the bootstrap samples. Here, the labeling is trivial on the log$(z)$ log$(a)$-plane, as can be seen in Figs.~\ref{early_M2} and~\ref{early_M3}. The labeling showed here is determined for each bootstrap sample by simply labeling the states by increasing extracted energy/mass values.

\begin{figure*}
\begin{subfigure}[t]{0.49\linewidth}
\includegraphics[width=1\linewidth]{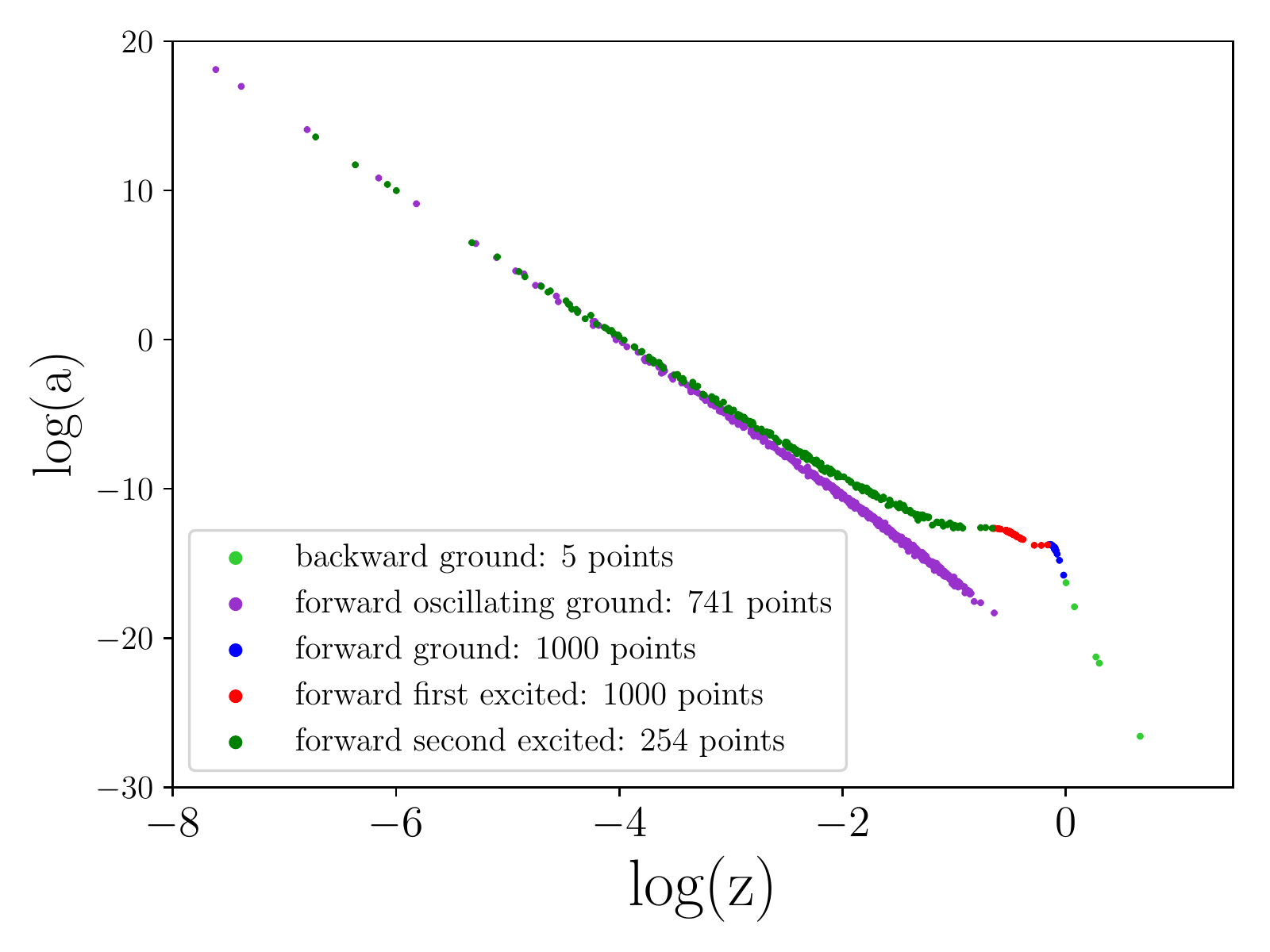}
\caption{Labeling states by type according to their mass.}
\label{M3_t5}
\end{subfigure}%
\hfill
\begin{subfigure}[t]{0.49\linewidth}
\includegraphics[width=1\linewidth]{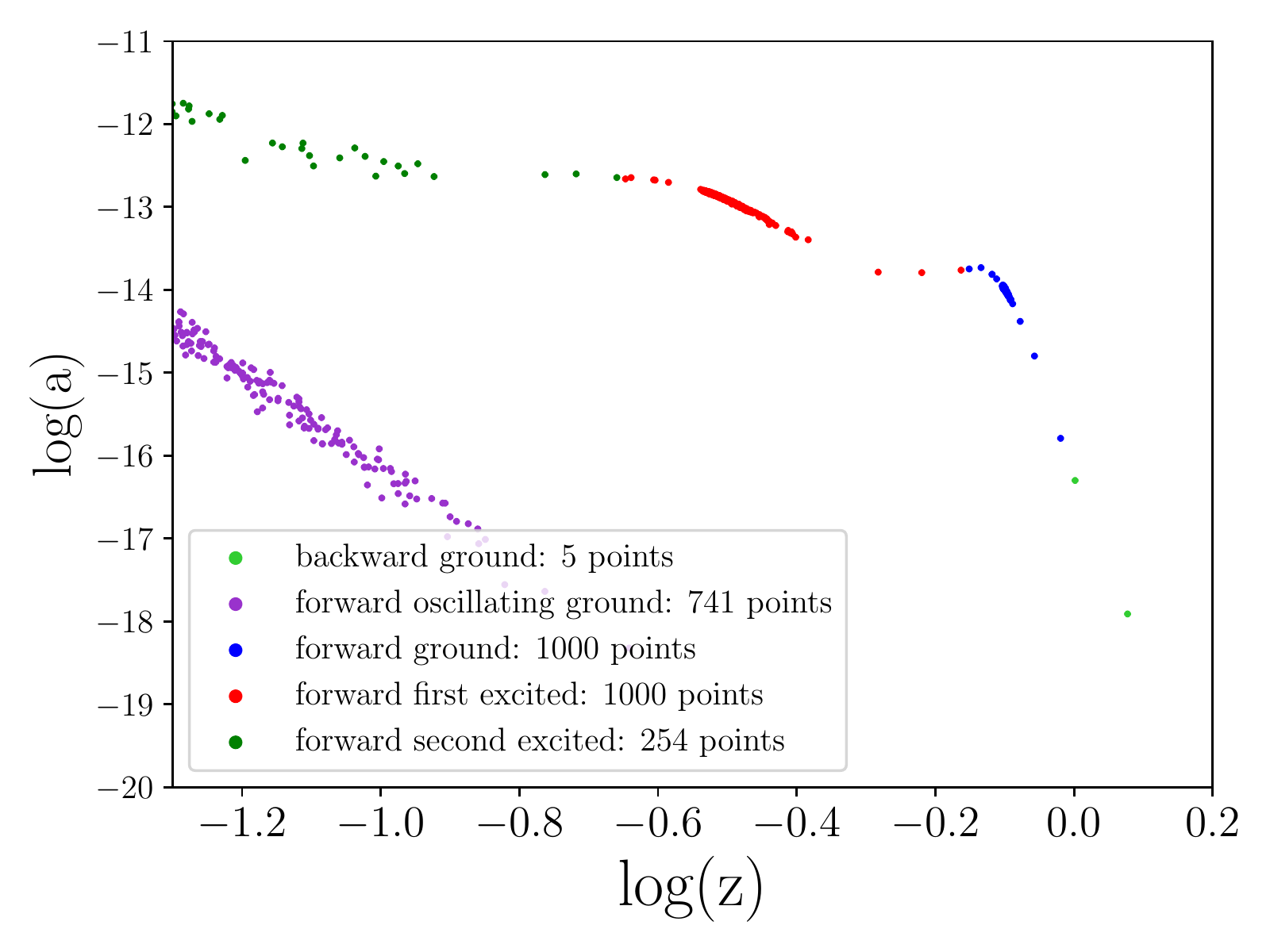}
\caption{Same as left, but zoomed into region where ground and first excited states lie.}
\label{M3_t5_zoomed}
\end{subfigure}%
\caption{Labeling according to mass when data is noisy, for $M=3$, $t=5$. Plot shows real part of logs.}
\label{M3_t5_both}
\end{figure*}

Unfortunately, the naive approach of labeling the states according to their extracted mass for each bootstrap sample breaks down for larger timeslices, where the variance of the clusters grows with noise on the stencil $y_n(t)$.  Figs.~\ref{M3_t5} and~\ref{M3_t5_zoomed} show an example of a situation where the naive cluster assignment based on the mass is not necessarily reliable. In this case, we can see more than three clusters appearing because for some bootstrap samples, one of the roots of the polynomial was $z_m > 1$, corresponding to $E_m < 0$. These states are identified in the figure as ``backwards propagating" states, although they may also represent downward statistical fluctuations on the ground state. There are also instances where one of the roots of the polynomial was $z_m < 0$, which are labeled as ``oscillating states." Here, there are clearly overlapping distributions, and finding the most probable state assignment is required to properly recover the central values and errors on the masses and amplitudes.

%
%
\section{Automated label flows}
In order to assign each bootstrap sample the correct set of labels, one may consider a clustering algorithm such as K-means~\cite{k_means}. However, points extracted from Prony's method have an additional constraint that each bootstrap sample of $M$ points contains signal from $M$ different distributions. An expectation maximization algorithm was proposed in Ref.~\cite{2019arXiv190210695C} which involved finding the most probable permutation of the $M$ points among $M$ clusters for each bootstrap sample. However, to compute the probabilities, one must assume a model of the probability distribution such as bivariate Gaussian distributed clusters, or use a non-parametric probability distribution/distance metric. Although a clustering method may be possible, we observe the failure of each of these methods for highly noisy, overlapping distributions as in Fig.~\ref{M3_t5_both}.

Instead of clustering the points as they appear directly from Prony's method, we propose an alternative approach, which we call {\it automated label flows}. The genesis of this idea stemmed from a discussion with Dan Hoying of RBC and UKQCD, where he described rescaling of errors, which is mentioned in his thesis (Ref.~\cite{Hoying}). We expand upon this method by introducing a robust approach to scaling errors back, such that no assumptions must made about the function which transforms the raw correlator data into a set of masses and matrix elements. The method of automated label flows is therefore applicable to any algebraic method of extracting excited states.

\subsection{Rescaling errors}
\begin{figure*}
\begin{subfigure}[t]{0.49\linewidth}
\includegraphics[width=1\linewidth]{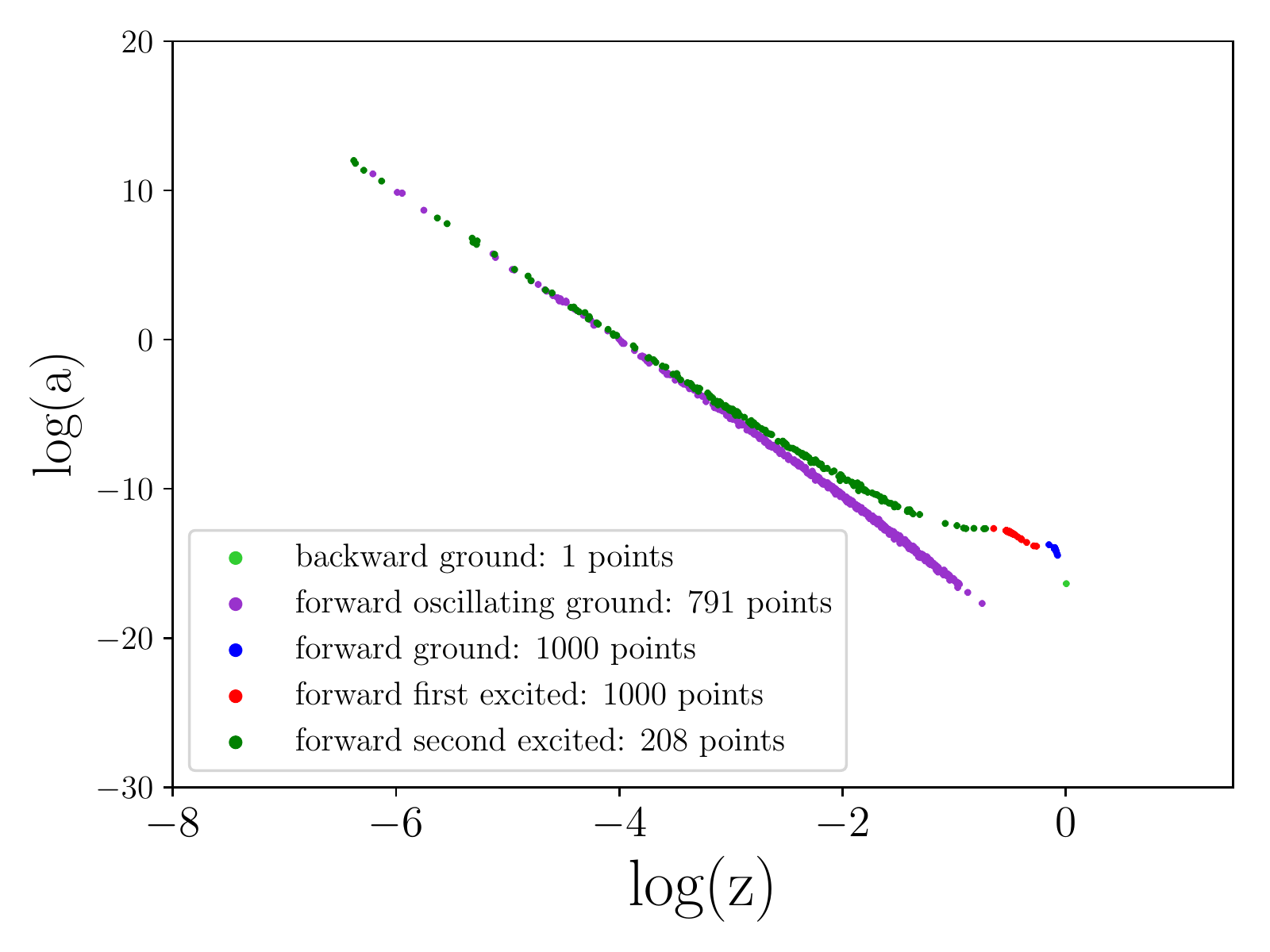}
\caption{Rescaled errors with $\epsilon = 0.8$.}
\label{M3_t5_e.8}
\end{subfigure}%
\hfill
\begin{subfigure}[t]{0.49\linewidth}
\includegraphics[width=1\linewidth]{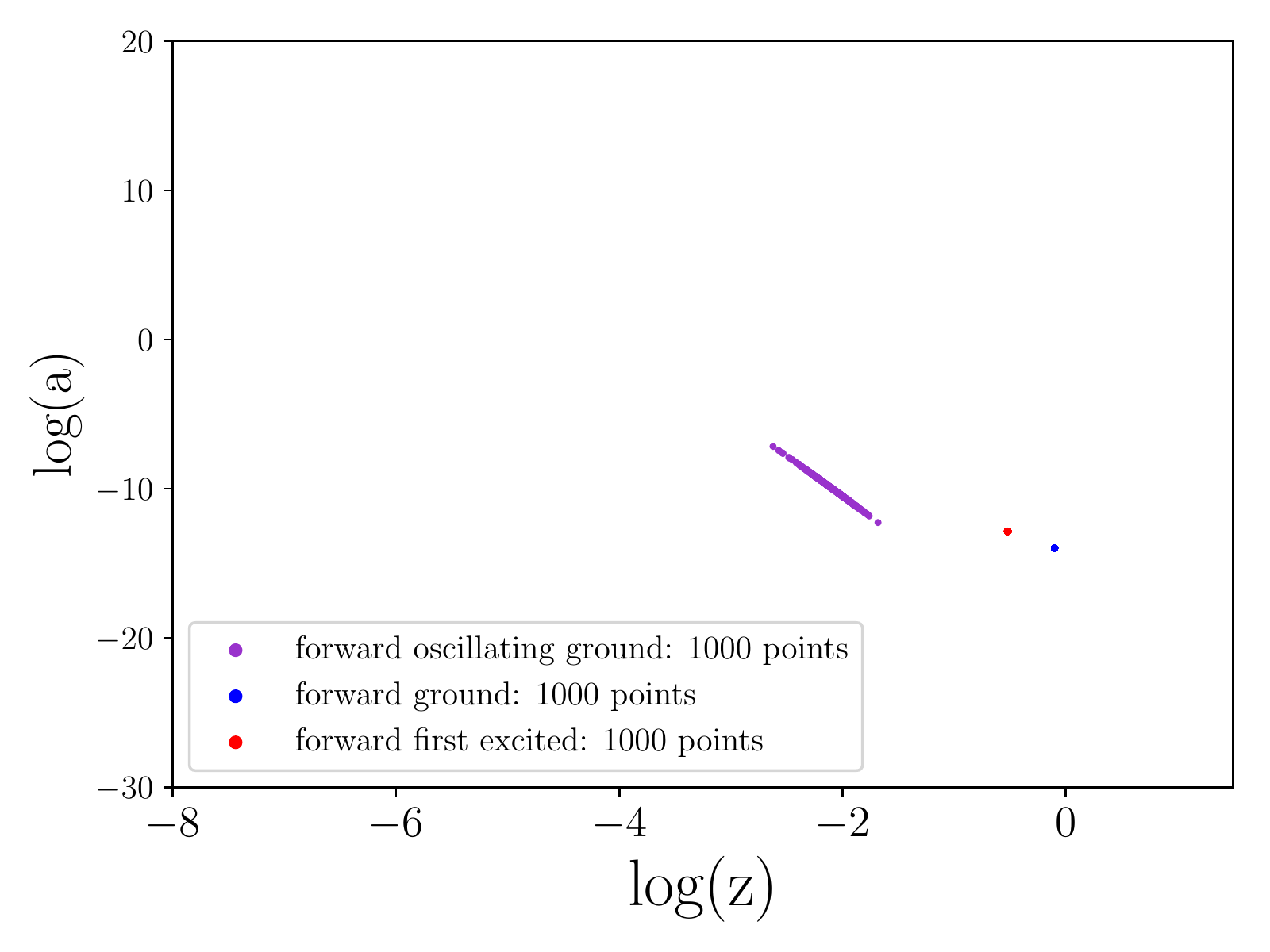}
\caption{Rescaled errors with $\epsilon = 0.1$.}
\label{M3_t5_e.1}
\end{subfigure}%
\caption{Rescaled errors for $M=3$ at $t=5$ clusters. Original clusters plotted in Fig.~\ref{M3_t5}.}
\label{M3_t5_scaled}
\end{figure*}

Our inability to label the states is a result of the large variance of the bootstrap samples of $y_n(t)$ which are fed into Prony's method. If the distribution of each timeslice in the stencil $y_n(t)$ was much more sharply peaked, the clusters would be more distinctly separated, and labeling would be trivial as it is in Figs.~\ref{early_M2} and~\ref{early_M3}. Consider a set of bootstrap samples, $\{x^{(i)}\}$, from some dataset with a variance of the bootstrap samples given by $\sigma_x^2$. Any given bootstrap sample, $x^{(i)}$, can be expressed as a deviation, $\delta x^{(i)}$, from the mean value of the dataset, $\bar{x}$ via
\begin{align}
x^{(i)} &= \bar{x} + \delta x^{(i)}. 
\end{align}
Then define 
\begin{align}
x^{(i)}_\epsilon &= \bar{x} + \epsilon \delta x^{(i)}.\label{epsilon}
\end{align}
For $\epsilon>0$, the distribution of $\{x^{(i)}_\epsilon\}$ has the same information as the original distribution of $\{x^{(i)}\}$, but a new variance given by  $\sigma_{x_\epsilon}^2 = \epsilon^2 \sigma_x$. Figs.~\ref{M3_t5_e.8} and~\ref{M3_t5_e.1} show the clusters obtained from Prony's method when we use $\epsilon = 0.8$ and $\epsilon = 0.1$, respectively, when $y_{\epsilon,\, n}^{(i)}(t)$ replaces the unscaled stencil $y_n^{(i)}(t)$, defined as in Eq.~\ref{epsilon}. In general, we find that for any dataset, choice of stencil, and model of $M$ states, we can always find a sufficiently small $\epsilon$ where the labeling is trivial, as in Fig.~\ref{M3_t5_e.1}.

\subsection{Automated flow for increasing $\epsilon$}
We propose to label the states iteratively in a series of $K$ steps from a small value where the labeling is trivial, $\epsilon_0$, to the original data where $\epsilon = 1$. In this dataset we find that $\epsilon_0 = 0.01$ is sufficiently small so that the cluster labeling is unambiguous. The series of values of $\epsilon$ can be indexed by $k$ taking integer values from 0 to $K$, so that 
 \begin{align} 
\epsilon_k \stackrel{\rm flow}{\longrightarrow} \epsilon_{k+1} &= \epsilon_k + \Delta \epsilon,
\end{align}
where we incrementally increase epsilon by 
\begin{align}
\Delta \epsilon = \frac{1- \epsilon_0}{K}. 
\end{align}
The flow will be determined by increasing $\epsilon$, 
\begin{align}
\epsilon_0\,  \stackrel{\rm flow}{\longrightarrow}\, \epsilon_1 \, \stackrel{\rm flow}{\longrightarrow}\, \cdots \,\stackrel{\rm flow}{\longrightarrow}\, \epsilon_{K-1} \,\stackrel{\rm flow}{\longrightarrow}\, \epsilon_K,
\end{align}
thus flowing the labels to the clusters of the original dataset ($\epsilon_K = 1$). In the results shown below, we use $\Delta \epsilon = 0.01$. \\

For each bootstrap sample, the movement of the $M$ points extracted from Prony's method are tracked for increasing $\epsilon$. For every iteration of the flow $k \rightarrow k+1$, a choice must be made to determine which of the $M$ points flowed from each of the $M$ points from the previous set. We choose that the label flow should be determined by selecting the permutation of the $M$ state labels that minimizes the distance traveled from the previous value of $\epsilon$ in the series.  That is, for each bootstrap sample, we find the permutation of labels, $\sigma = (1,2, 3, ...), (2,1,3,...)$ or $(3,1,2,...)$, etc. which minimizes the sum over the $M$ distances $d(m; \sigma_m)$, 
\begin{align}
\argmin_\sigma \sum_{m = 1}^M d(m; \sigma_m),\label{minimize}
\end{align}
where $d(m; \sigma_m)$ is the distance between the $m^{\rm th}$ point from the extraction with $\epsilon = \epsilon_k$ and the $\sigma_m\,^{\rm th}$ point from the extraction with $\epsilon = \epsilon_{k+1}$. We define this distance to be
\begin{widetext}
\begin{align}
d(m;\sigma_m) &= \sqrt{ \Bigg|\frac{\log(z_m^{(\epsilon_k)}) - \log(z_{\sigma_m}^{(\epsilon_{k+1})})}{\Delta \log(z)}\Bigg|^2 + \Bigg|\frac{\log(a_m^{(\epsilon_k)}) - \log(a_{\sigma_m}^{(\epsilon_{k+1})})}{\Delta \log(a)}\Bigg|^2}, \label{metric}
\end{align}
\end{widetext}
where we use log space because the clusters appear to be more elliptical in shape in log space, lending themselves to the Euclidean-type metric used here. We also account for the large separation of scales between the masses and matrix elements by normalizing the scale difference between $\log(z)$ and $\log(a)$, normalized by the total extent of the clusters in those directions, $\Delta \log(z)$ and $\Delta \log(a)$. This normalization is necessary because, for example in this dataset, the log of the amplitudes, $a_m$, vary over a much larger range than the energies (log of the roots, $z_m$). For example, the clusters given from the $M=3, t = 5$ extraction shown in Fig.~\ref{M3_t5_both} range in the $y$-direction from -29 to 19, whereas the $x$-direction ranges only from about 1 to -8. In other words, we use the scales of what would be the covariance matrix of a bivariate Gaussian fitted to the clusters to set the scales of the distance metric.  Note also that the roots and amplitudes are in general complex valued, so we compute the distance via the modulus squared. 

\begin{figure}
\centering
\includegraphics[width=1\linewidth]{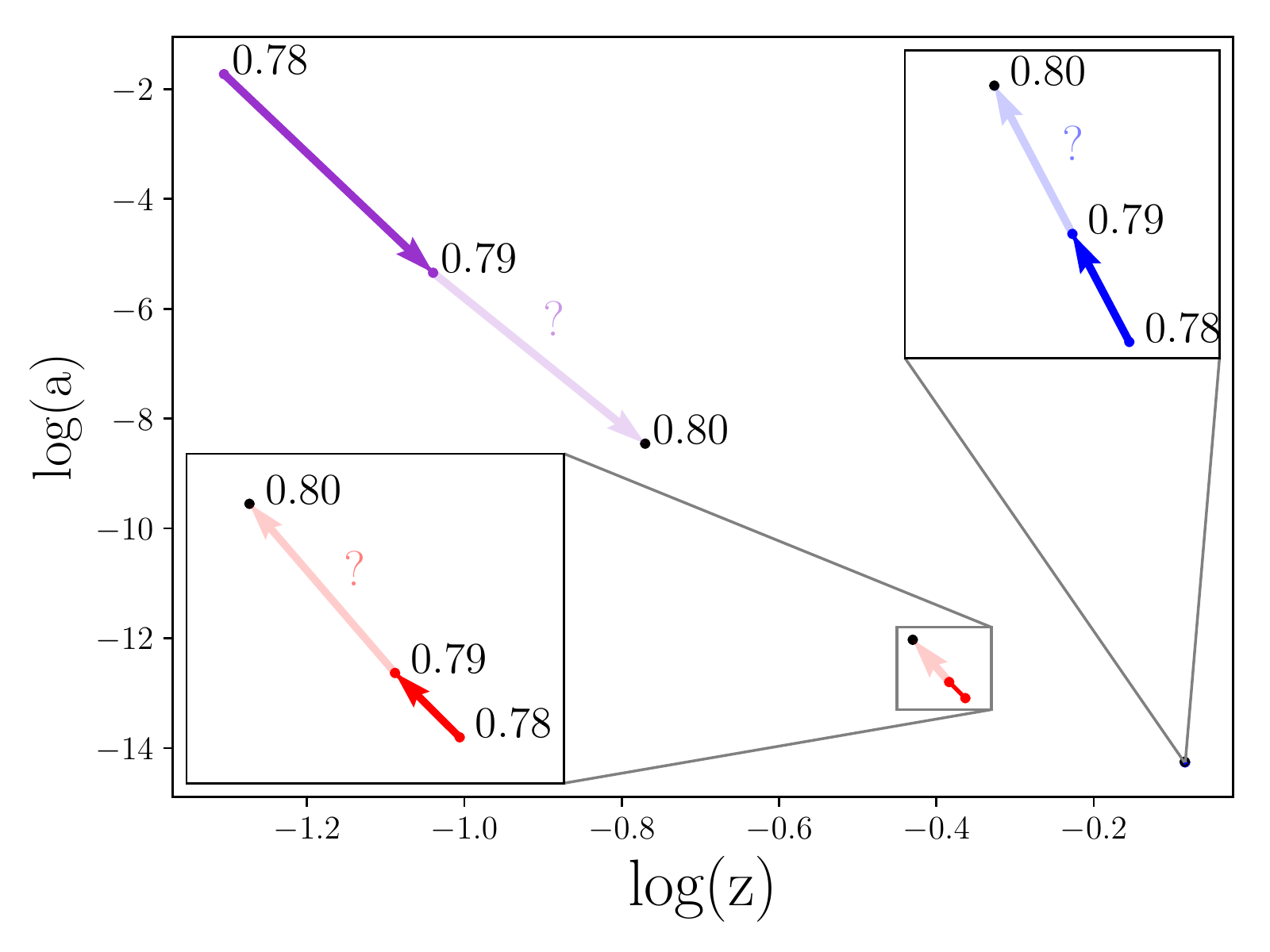}
\caption{Example of labels flowing from $\epsilon = 0.78$ to $\epsilon = 0.79$ to $\epsilon = 0.80$ for the $11^{\rm th}$ bootstrap sample of the $M=3$, $t=16$ extraction. The blue arrows, red arrows, and maroon arrows represent the flow of the ground state, first excited state, and second excited states, respectively.}
\label{example_flow}
\end{figure}

Fig.~\ref{example_flow} shows an example of label flows for the $11^{\rm th}$ bootstrap sample of the $M=3$, $t=16$ extraction. Shown in the figure are the three points $(z_m, a_m)$ on the $\log(a)$ $\log(z)$ plane for three values of $\epsilon$: 0.78, 0.79, and 0.80, as labeled in the figure. These are part of a flow with $\epsilon_0 = 0.01$, and $\Delta\epsilon= 0.01$ ($\epsilon_{77} = 0.78, \epsilon_{78} = 0.79, \epsilon_{79} = 0.80$). The label flow has been determined for $\epsilon_{77}\,  \stackrel{\rm flow}{\longrightarrow}\, \epsilon_{78}$ as indicated by the solid blue (ground state), red (first excited state), and maroon (second excited state) arrows. The points corresponding to $\epsilon_{79} = 0.80$ are colored in black to represent the ambiguity present in the labels before a choice of labels is selected. Upon executing the labels flow from $\epsilon_{78}$ to $\epsilon_{79}$ according to Equations~\ref{minimize} and ~\ref{metric}, the labels will flow according to the faded arrows.

\begin{figure*}
\begin{subfigure}[t]{0.49\linewidth}
\includegraphics[width=1\linewidth]{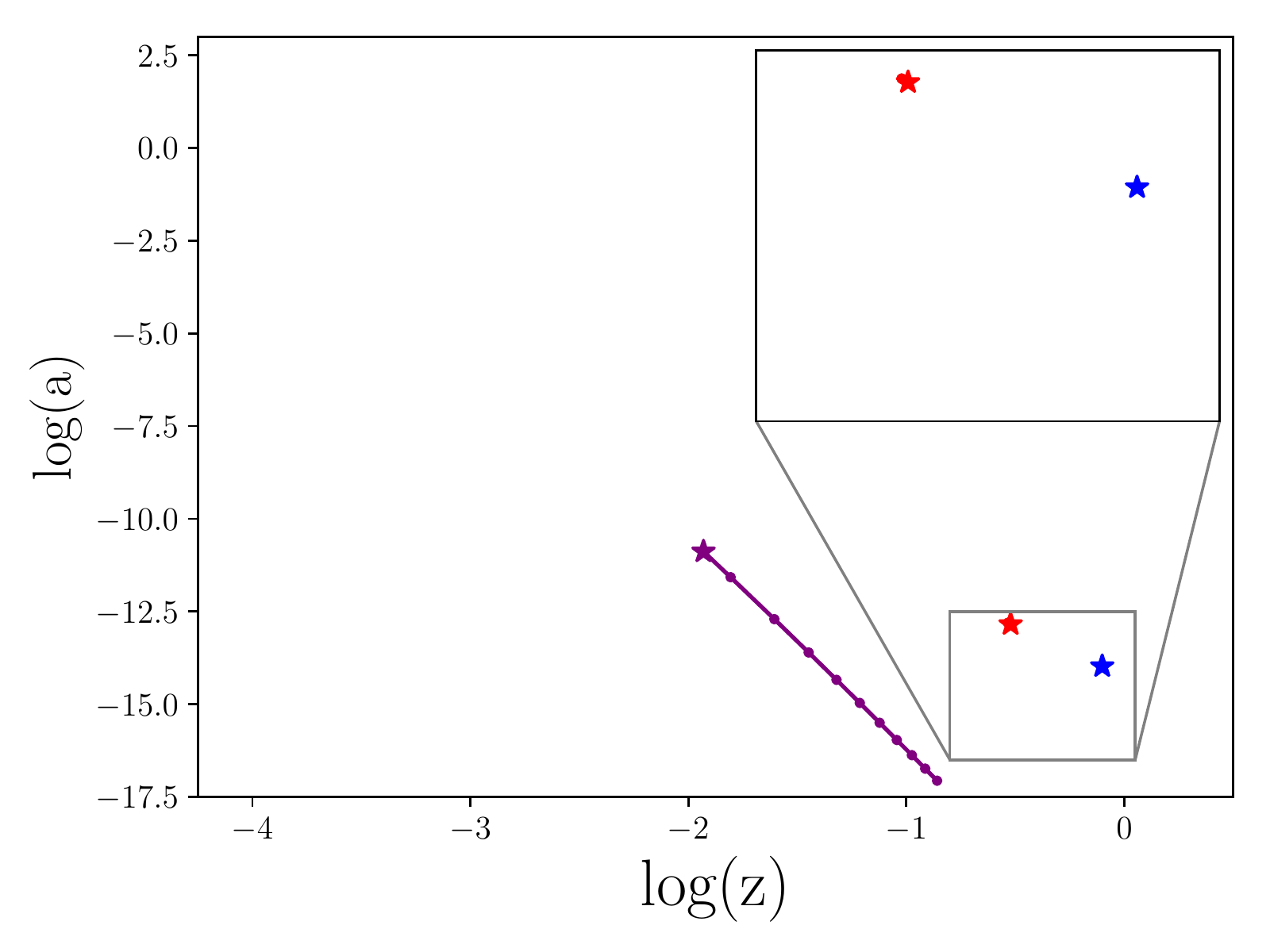}
\caption{Flow of labels for first bootstrap sample.}
\label{flow_1}
\end{subfigure}%
\hfill
\begin{subfigure}[t]{0.49\linewidth}
\includegraphics[width=1\linewidth]{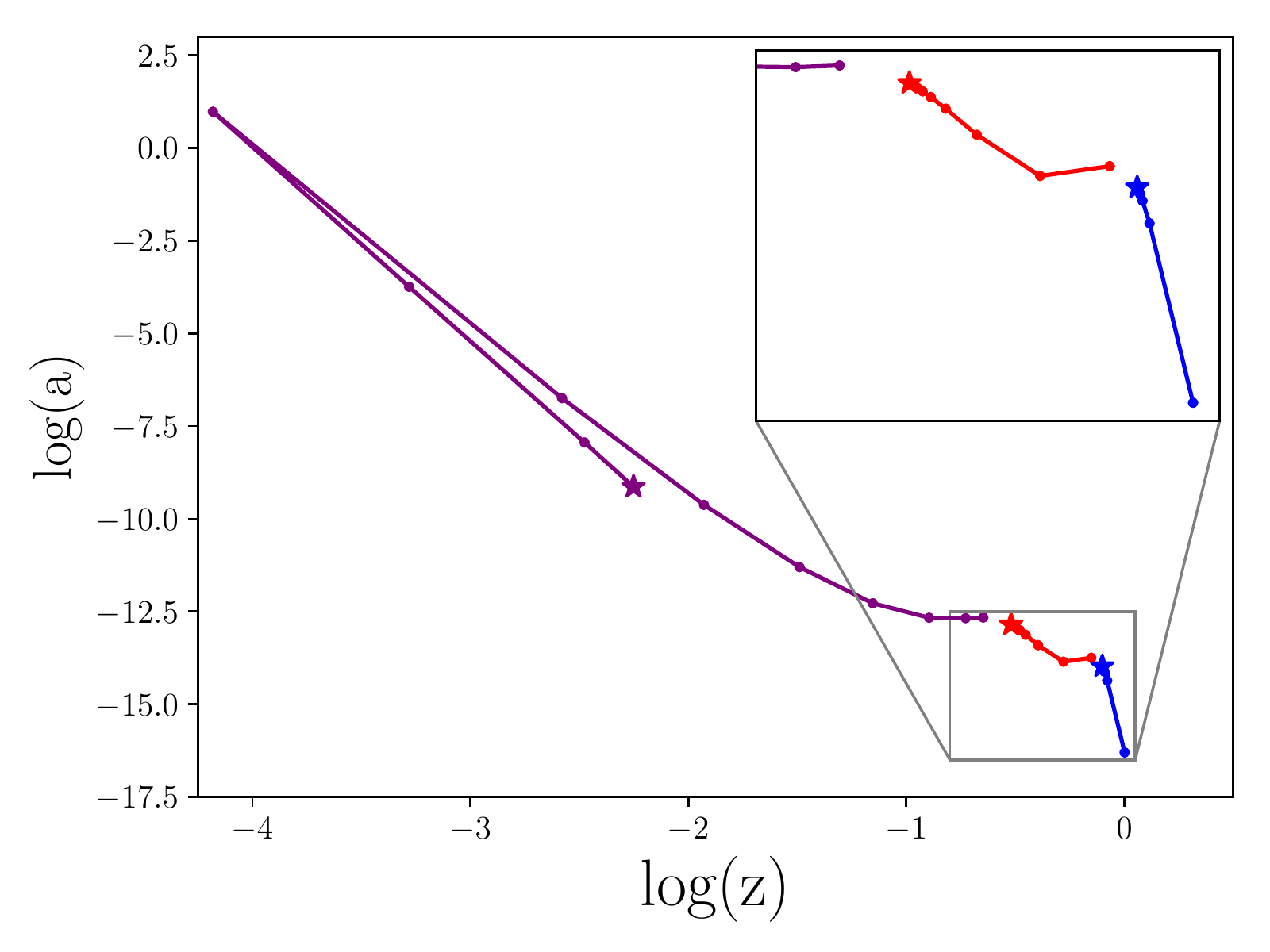}
\caption{Flow of labels for 33rd bootstrap sample.}
\label{flow_2}
\end{subfigure}%
\caption{Label flows for two different bootstrap samples of $M=3$ at $t=5$ clusters, starting from $\epsilon = 0.05$, denoted by the starred point, then flowing through $\epsilon = 0.1, 0.2, 0.3, \ldots, 0.9,$ to $\epsilon = 1.0$  }
\label{flow}
\end{figure*}

\begin{figure*}
\begin{subfigure}[t]{0.49\linewidth}
\includegraphics[width=1\linewidth]{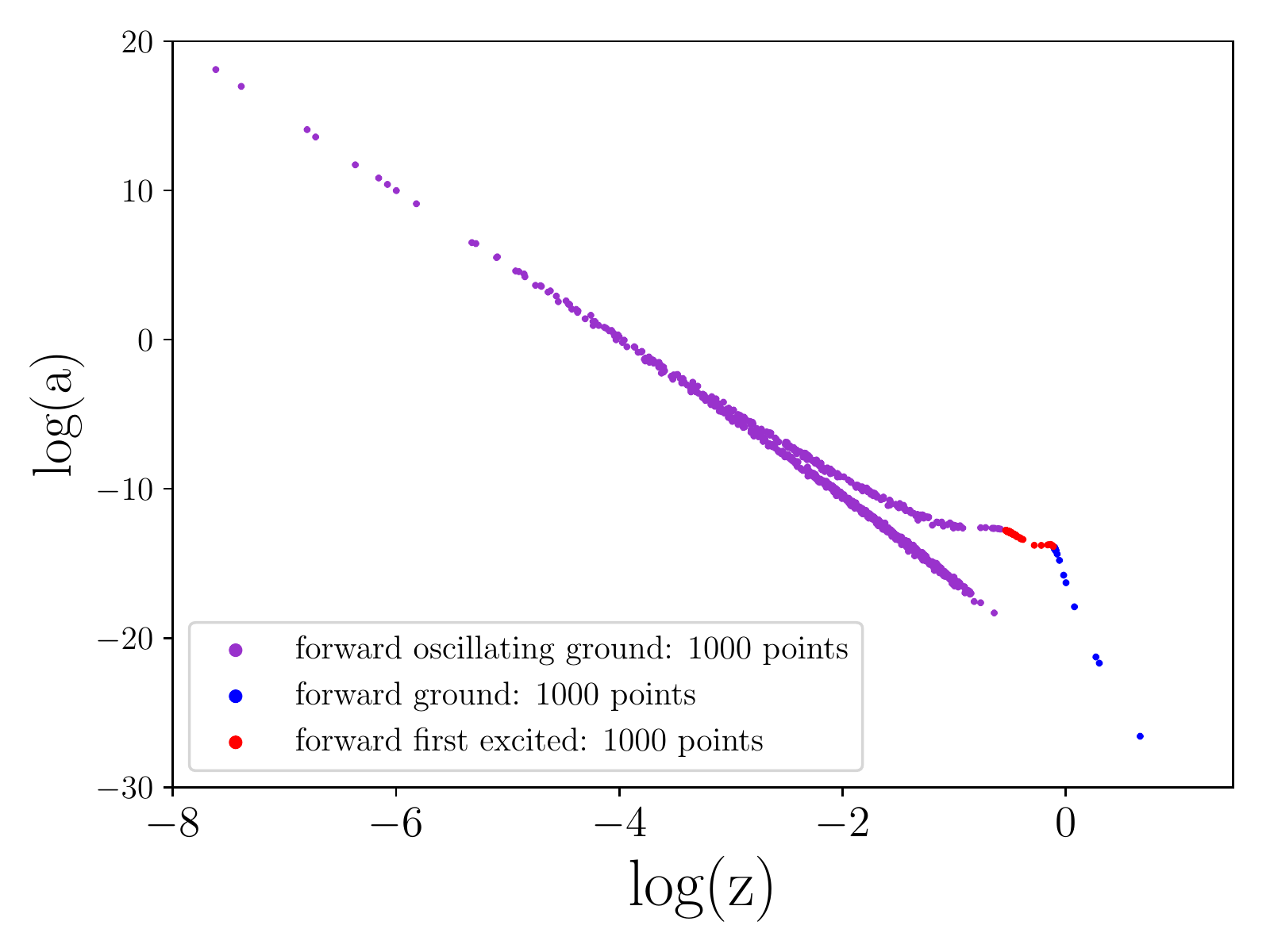}
\caption{Flowed clusters.}
\label{flowed_M3_t5_unzoomed}
\end{subfigure}%
\hfill
\begin{subfigure}[t]{0.49\linewidth}
\includegraphics[width=1\linewidth]{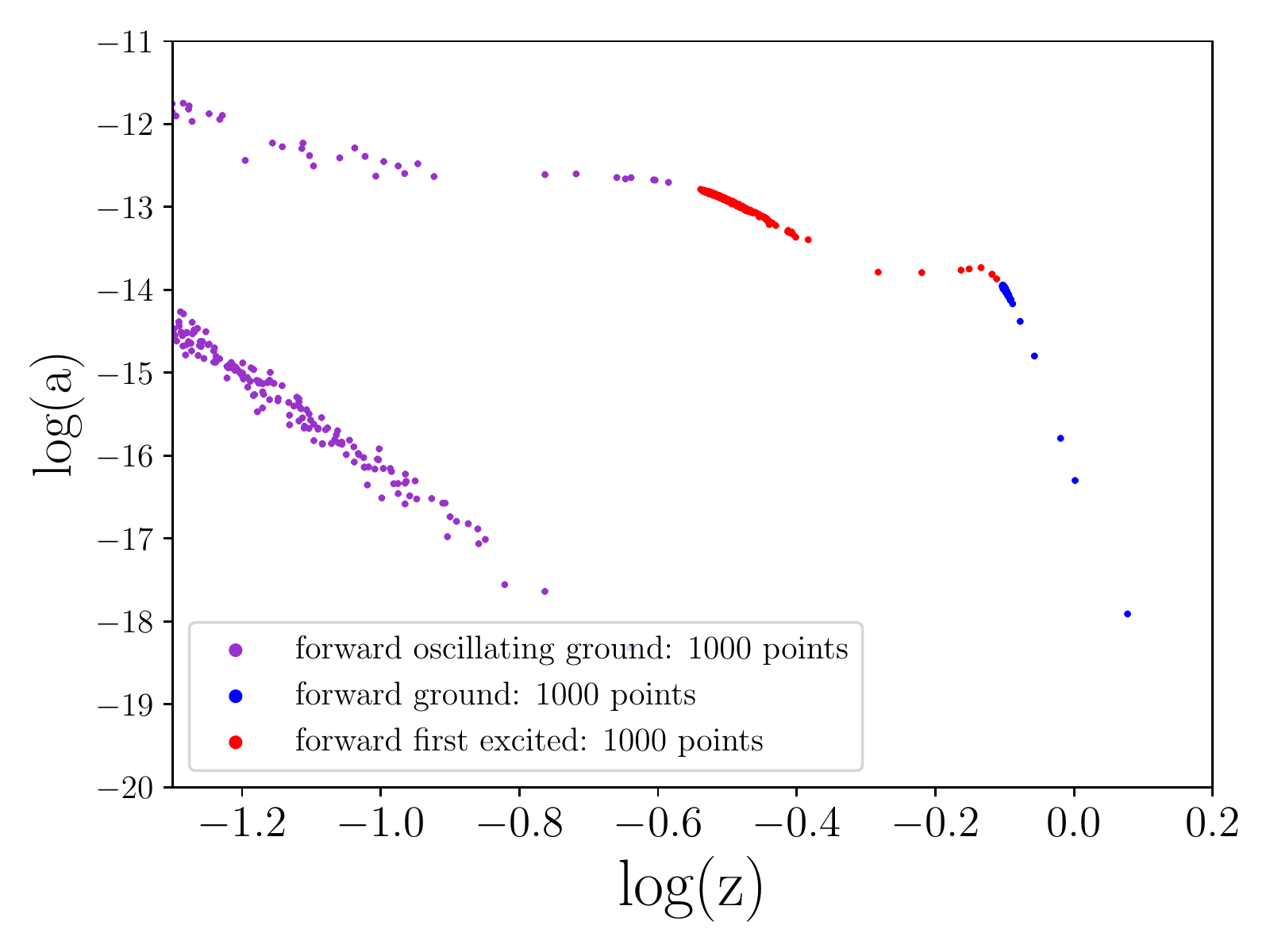}
\caption{Same as left, but zoomed into region where ground state and first excited state lie.}
\label{flowed_M3_t5_zoomed}
\end{subfigure}%
\caption{Clusters with flowed labels for $M=5$ at $t=5$.}
\label{flowed}
\end{figure*}

An example of a full label flow for two bootstrap samples can be seen in Fig.~\ref{flow}. In the figure, the label flow begins at $\epsilon_0$ indicated by the starred point. The resulting flowed clusters for $\epsilon = 1.00$ can then be used to estimate the errors on the extracted masses and amplitudes. Flowing each of the bootstrap samples in this way determines the labels of the clusters, an example of which can be seen in Fig.~\ref{flowed}.

Also note that in each of these example figures, we interpret the points that were previously identified as backwards propagating by the naive cluster assignment to actually be downward statistical fluctuations on the ground state resulting in a negative. This way, only if the mean of a cluster is backwards propagating is a point identified as backwards propagating. Although this choice is not the only possibility, we make the choice to follow the flow from its mean value to ensure a deterministic flow. Applying this across all types of extracted states, every stencil used in a model with $M$ states yields $M$ clusters.

 
\section{Discussion and Results}
In this section, we make note of some interesting behaviors of the flows, and describe how we deal with them. We then compile the resulting spectrum results and compare them to those obtained from a chi-squared minimization fit to a sum of exponentials. 

\subsection{Discussion of abnormal label flows}
We note that for some timeslices, the label assignments may be ``incorrect." For example, in Figs.~\ref{flowed_M3_t5_unzoomed} and~\ref{flowed_M3_t5_zoomed}, some of the points labeled ``forward oscillating" may actually be signal from the second excited state (compare to Fig.~\ref{M3_t5}). This situation occurs when the mean of the stencil $y_n(t)$ corresponds to a different state than that of the particular bootstrap sample. An example of this type of flow can be seen in Fig.~\ref{flow_2}, where arrows connect the points for each state, representing the direction of increasing $\epsilon$. Here, the maroon state begins as an oscillating state, and follows a very long trajectory towards high masses, and then rapidly changes directions towards lower masses. At the cusp of its trajectory, the root $z_m$ changes sign, from negative (oscillating state) to positive (second excited state). However, since the label as an oscillating state is assigned at $\epsilon = 0.01$, the state that flows to $\epsilon = 1.00$ is identified as an oscillating state. This behavior likely stems from fact that we are choosing a value of $M$ too small, where there is actually signal for more than $M$ states. Therefore, it is possible that this problem may be minimized by increasing $M$. In any case, for a given $M$, our algorithm is designed to identify the $M$ states with the largest signal in the {\it average} ensemble, whereas any particular bootstrap sample, or a different ensemble entirely, will have its own unique signal, and thus may give a different set of $M$ states from Prony's method.  

In most cases, we do not find such mislabeling of states to be a problem because the clusters of such states are extremely large in comparison with the states where this problem does not occur. That is, an incorrect labeling is inconsequential for all practical purposes because the error on the mass and matrix elements would be too large to obtain any meaningful physics. We choose a trade off and prefer increasing inaccuracy in noisy states to obtain higher accuracy in lower lying and less noisy states. This choice is made because it provides a unique criteria for labeling, though it is not necessarily the {\it best} choice. It is not clear that there {\it is} a correct, or best, criteria for labeling, so we are simply attempting to make a deterministic choice. 

\begin{figure*}
        \centering
        \begin{subfigure}[b]{0.49\textwidth}
         \centering 
        \includegraphics[width=1\textwidth]{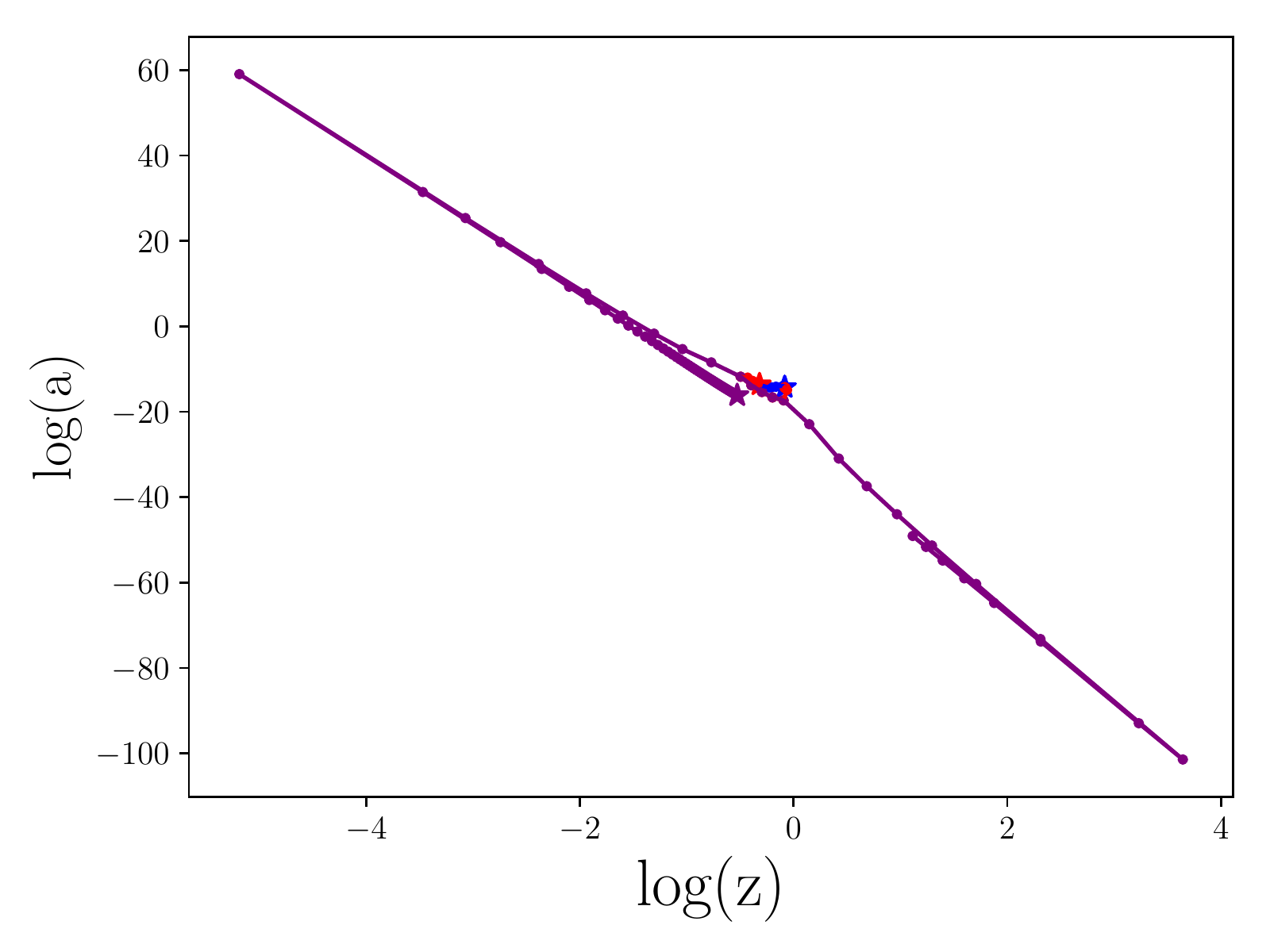}
        \caption{Flow depicted with arrows pointing in direction of increasing $\epsilon$.}
        \label{crazy_unzoomed}
         \end{subfigure}%
        \hfill
         \begin{subfigure}[b]{0.49\textwidth}
         \centering 
        \includegraphics[width=1\textwidth]{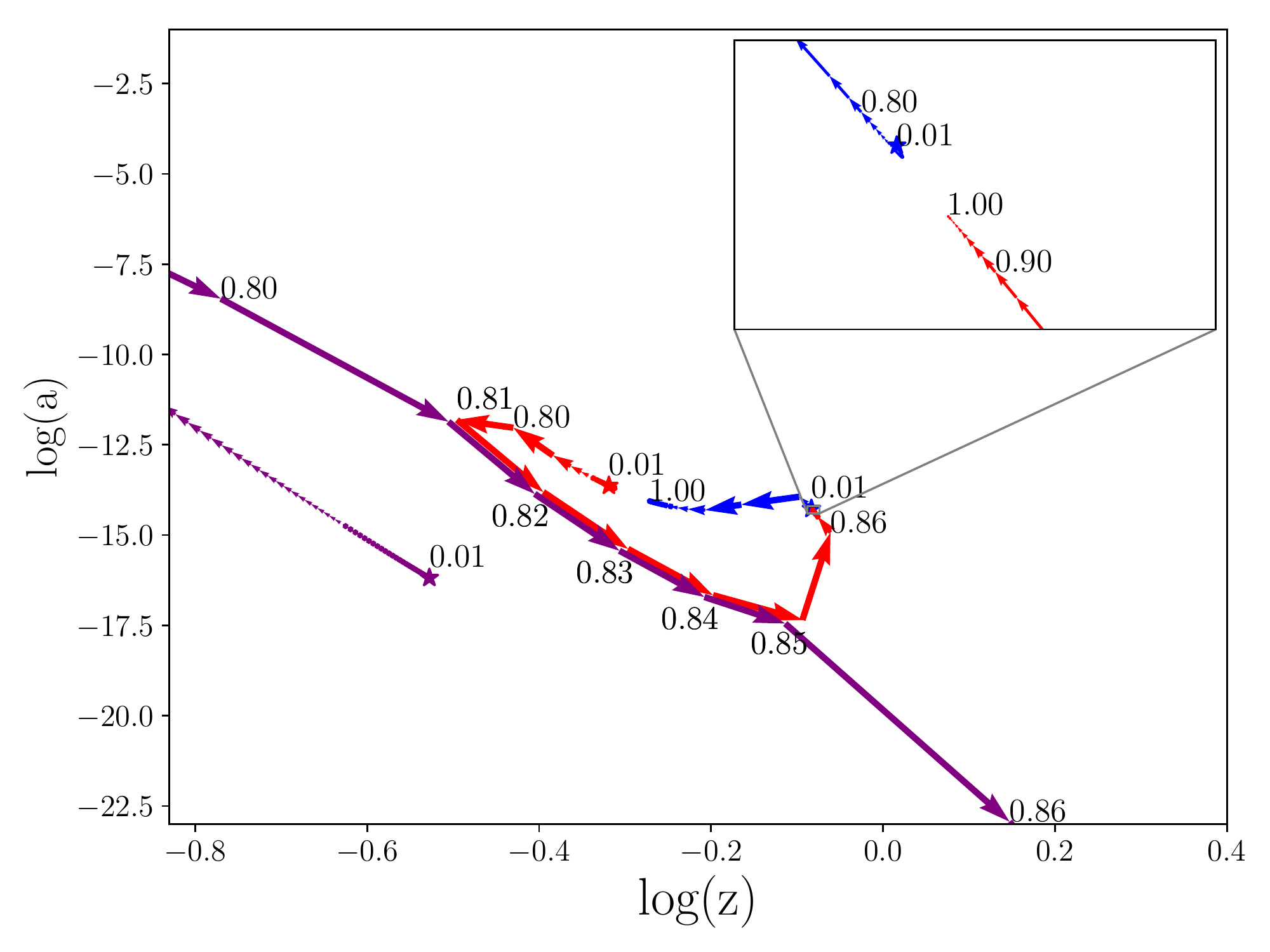}
        \caption{Label flow on left, zoomed into region where the flow is ambiguous.}
        \label{crazy_zoomed}
         \end{subfigure}%
         \caption{Label flow for $11^{\rm th}$ bootstrap sample of $M=3, t=16$ extraction, exhibiting a label collision.}
        \label{crazy}
    \end{figure*}

In a small fraction of bootstrap samples ($<$1\% for this data set), even the automated label flows method does not provide a completely deterministic labeling of states. Fig.~\ref{crazy} depicts an example from the $M=3$, $t = 16$ extraction (the same sample used in Fig.~\ref{example_flow}) where a {\it label collision} occurs at $\epsilon = 0.81$ and there is a {\it label merge} until $\epsilon = 0.85$. In this example bootstrap sample, a small $\Delta \epsilon$ is required to resolve the peculiar dynamics which have occurred. To track this behavior, the flow of $\epsilon$ is given by \{0.01, 0.02, ..., 0.99, 1.00\}. As seen in Fig.~\ref{crazy_zoomed} the ground state (blue) follows a straightforward trajectory, but the first excited state (red) and second excited state/oscillating state (maroon) merge at $\epsilon = 0.81$ and separate again at $\epsilon = 0.85$. For $\epsilon = 0.81, 0.82, 0.83, 0.84, 0.85$, two of the roots extracted from Prony's method are complex conjugate pairs, so the distance between points of increasing epsilon are identical. Whenever there are complex roots for a given value of $\epsilon$ for a given bootstrap sample, a collision is inevitable, even if the average ensemble does not have degeneracy in its spectrum. Whenever this happens, the merging of states represents an ambiguity in the flow because once the states separate, a choice must be made as to which label flows along which trajectory.

One way to address this ambiguity is to apply each permutation of the collided states with equal probability. Another way could be to choose the permutation based on the history of the flow. In the case depicted here, the second excited/oscillating state follows a very long trajectory before the collision compared to the ground state and first excited state. Based on this history, the state with the longer trajectory after the label merge could be attributed to the second excited/oscillating state. If this permutation is not chosen, the first excited state may flow to a large negative mass, and the ``incorrect assignment" of this bootstrap sample would skew the variance on the average mass of the first excited state. Therefore, we choose to minimize the variance on the low lying states by choosing the permutation based on the history of the flow. Interestingly for the small subset of bootstrap samples that follow a flow as in Fig.~\ref{crazy}, the first excited state (based on trajectory length described above) has a mass smaller than the mass of the ground state due to the looping behavior as in Fig.~\ref{crazy_zoomed}. Hence, automated label flows results in a different labeling than the naive clustering based on the mass of the states.

\subsection{Effective mass plots and comparison to least squares fits}
\begin{figure*}
        \centering
        \begin{subfigure}[b]{0.49\textwidth}
            \centering
            \includegraphics[width=1\textwidth]{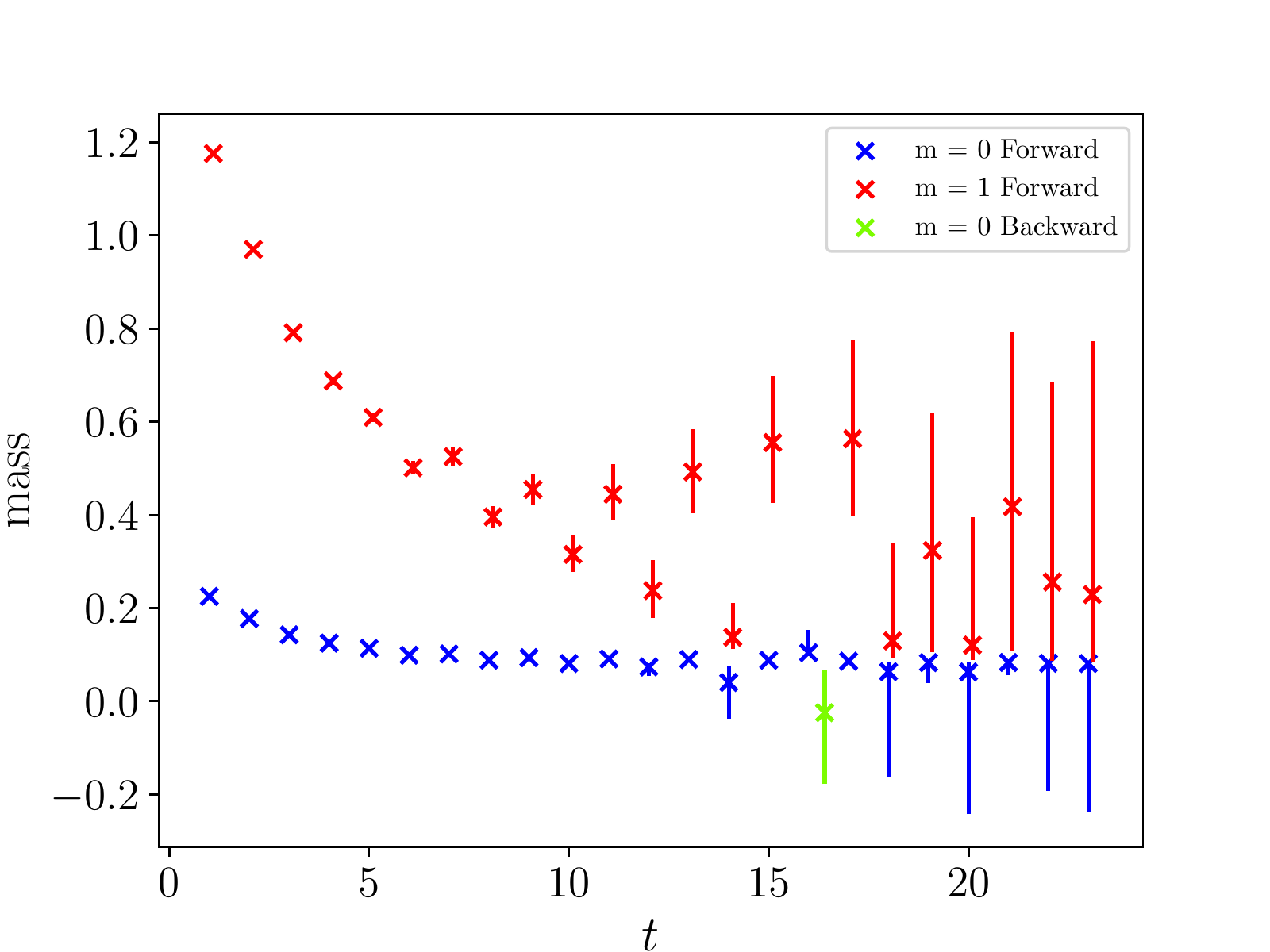}
            \caption{$M=2$ state extraction effective mass.}    
            \label{M2_effective}
        \end{subfigure}
        \hfill
        \begin{subfigure}[b]{0.49\textwidth}  
            \centering 
            \includegraphics[width=1\textwidth]{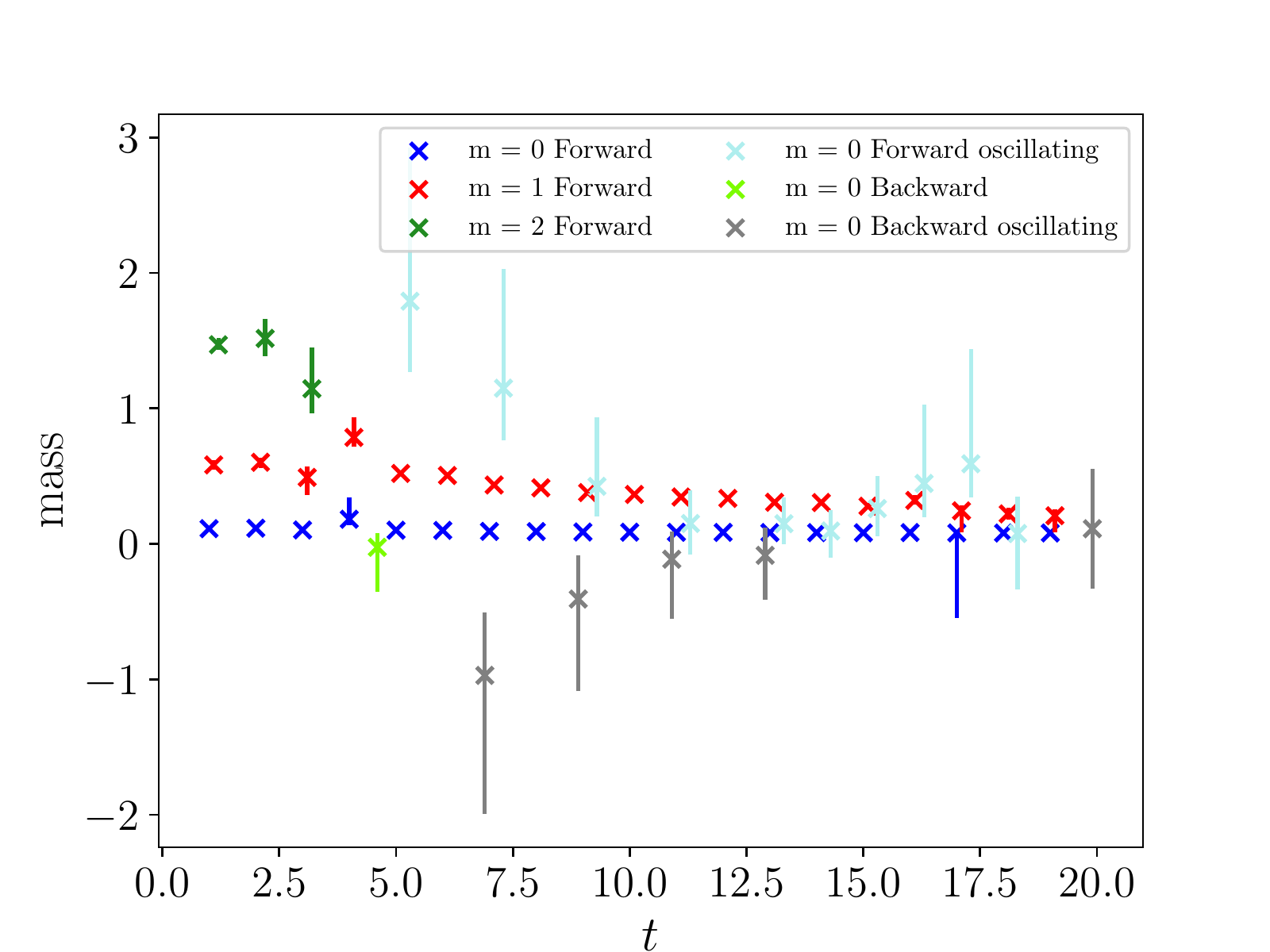}
            \caption{$M=3$ state extraction effective mass. }  
            \label{M3_effective}
        \end{subfigure}
        \vskip\baselineskip
        \begin{subfigure}[b]{0.49\textwidth}   
            \centering 
            \includegraphics[width=1\textwidth]{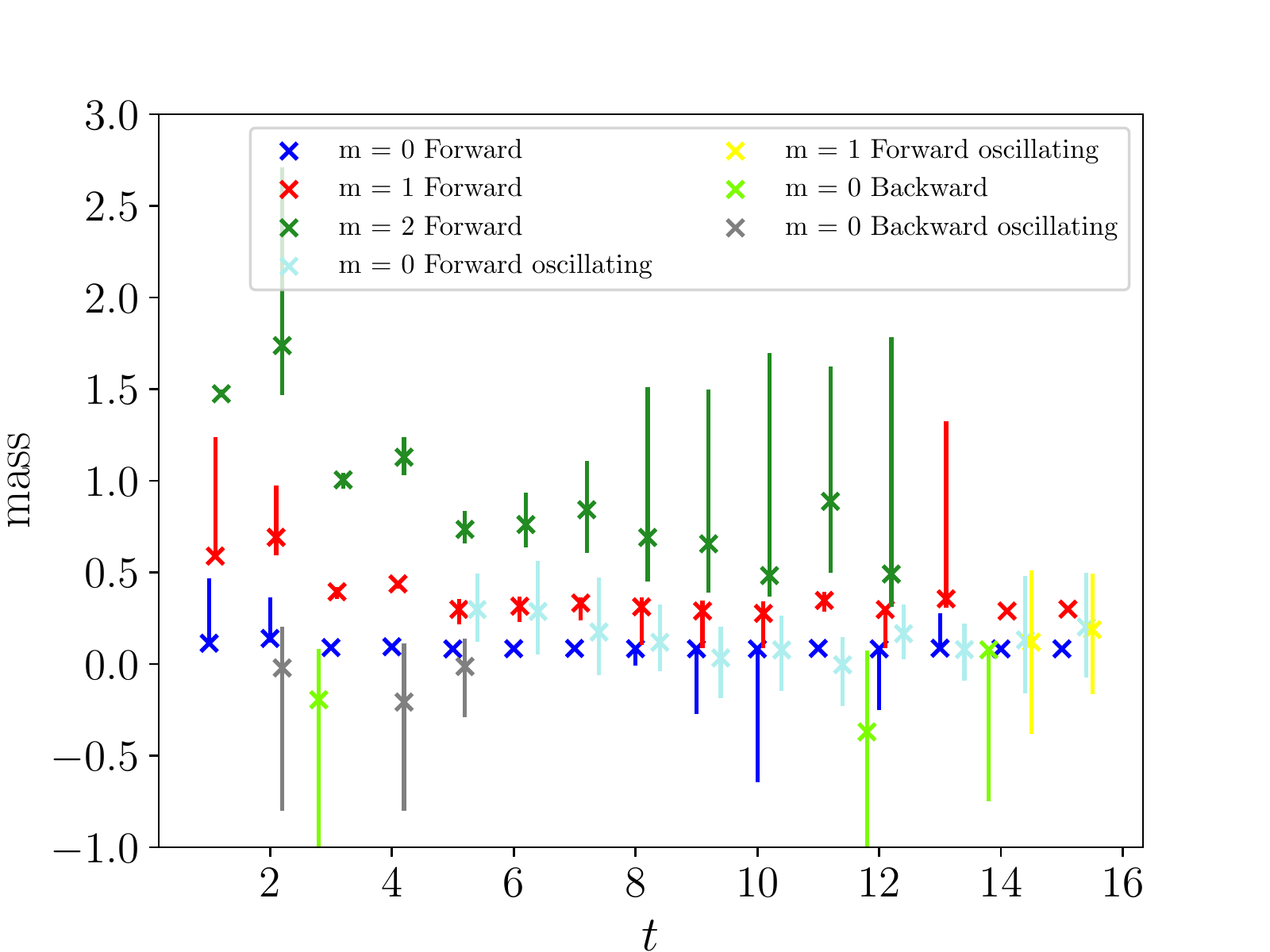}
            \caption{$M=4$ state extraction effective mass.}    
            \label{M4_effective}
        \end{subfigure}
        \hfill
        \begin{subfigure}[b]{0.49\textwidth}   
            \centering 
            \includegraphics[width=1\textwidth]{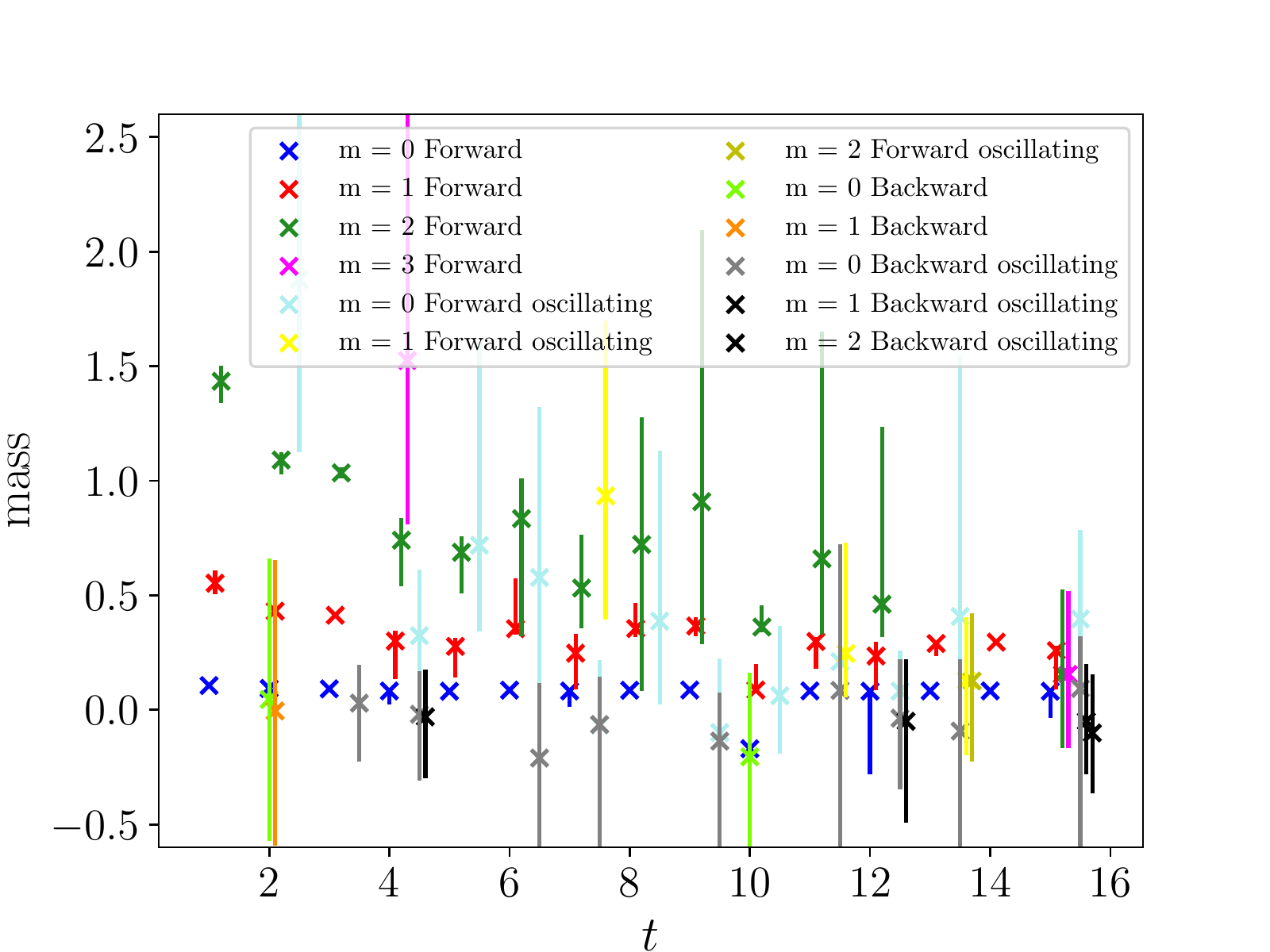}
            \caption{$M=5$ state extraction effective mass.}    
            \label{M5_effective}
        \end{subfigure}
        \caption{Generalized effective masses from Prony's method using automated label flows. The $x$-axis indicates the timeslice $t$ at the {\it beginning} of the Prony stencil of $2M$ timeslices, i.e. the timeslices $t$ from a given stencil $y_n(t)$.}
        \label{effective}
    \end{figure*}

Once a consistent set of labels is found, one can analyze generalized effective mass plots that are obtained from Prony's method. Since the state extractions in Prony's method depend on the particular stencil chosen, $y_n(t) = C(t+n)$, the states can be plotted as a function of Euclidean time, and an effective mass plateau can be found for each of the states that is extracted. Fig.~\ref{effective} shows the generalized effective mass plots of the $M=2$ through $M=5 $ state extractions from Prony's method. The clusters are in general not Gaussian distributed, so the error bars given in the effective mass plots correspond to the $16^{\rm th}$ and $84^{\rm th}$ percentiles around the median for each cluster at each timeslice, which would coincide with the $1 \sigma$ errors for Gaussian distributed data. In the plots below, we show all of the extracted states with their error bars obtained from automated label flows with $\epsilon = \{0.01, 0.02, ..., 0.99, 1.00\}$.

\begin{figure*}
\begin{subfigure}[t]{0.49\linewidth}
\includegraphics[width=1\linewidth]{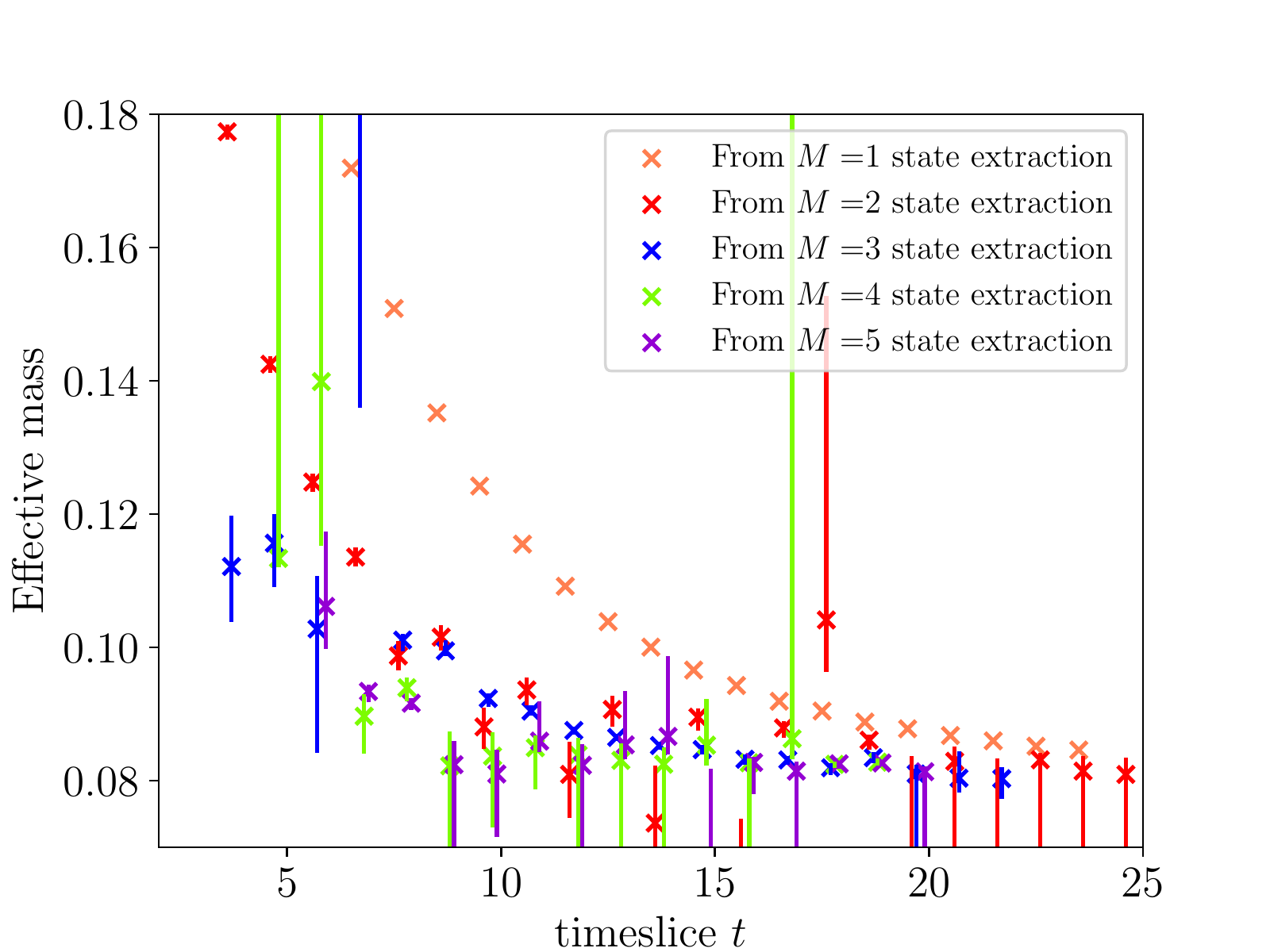}
\caption{Effective masses of ground state.}
\label{ground_compare}
\end{subfigure}%
\hfill
\begin{subfigure}[t]{0.49\linewidth}
\includegraphics[width=1\linewidth]{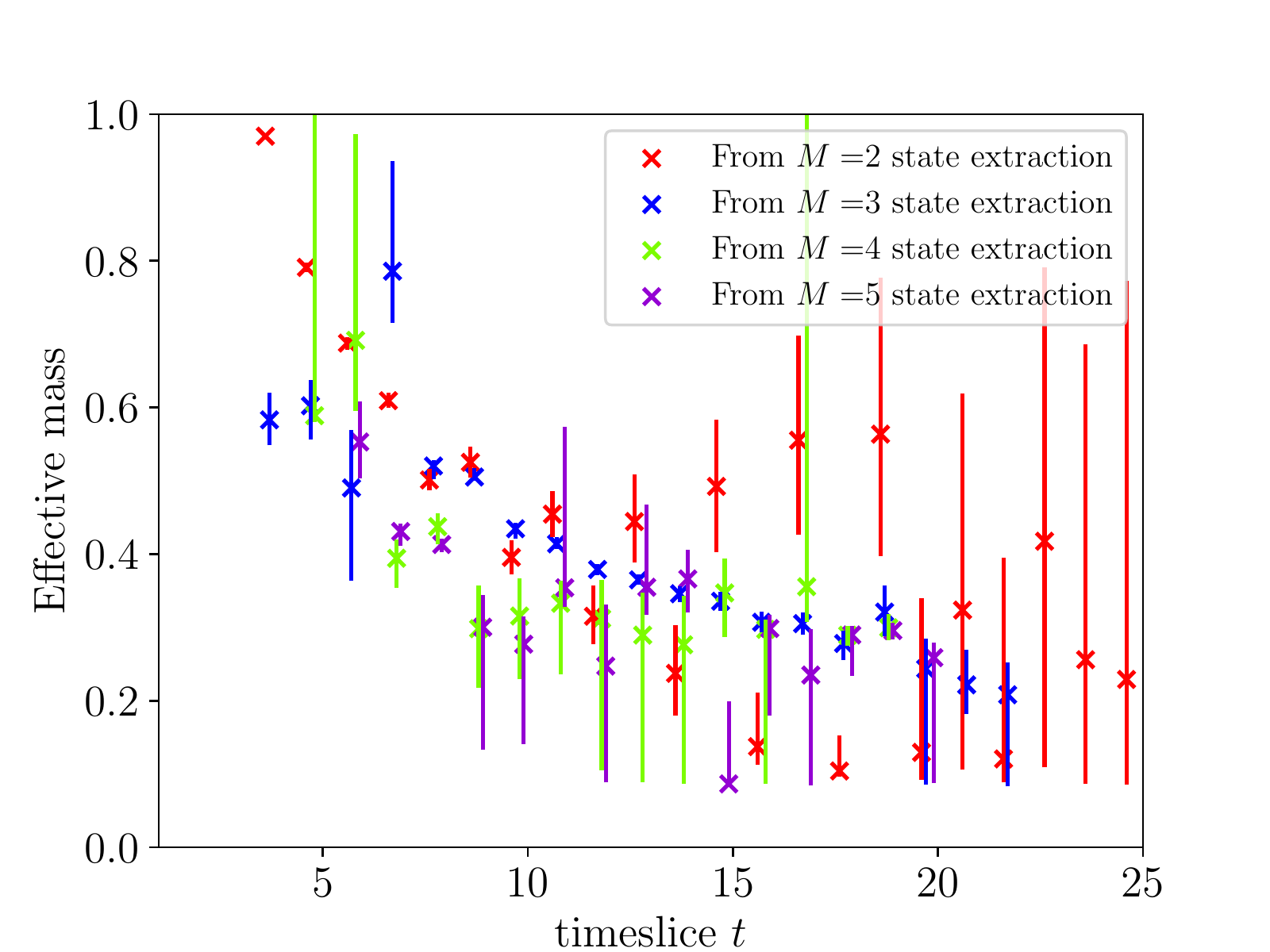}
\caption{Effective masses of first excited state.}
\label{first_compare}
\end{subfigure}%
\caption{Comparison between effective masses obtained from $M=1,2,3,4,5$ state extractions from Prony's method. For easy comparison between different numbers of states $M$, the $x$-axis indicates the {\it middle} timeslice used in the Prony stencil of $2M$ timeslices. }
\label{compare_effective}
\end{figure*}

It is interesting to observe how the effective masses of a state varies as a function of the number of states extracted, $M$. Fig.~\ref{ground_compare} compares the ground state effective mass for $M = 1,2,3,4,5$, and Fig.~\ref{first_compare} compares the first excited state effective masses for $M=2,3,4,5$. For fair comparison between models, the effective mass is plotted at the timeslice corresponding to the middle of the Prony stencil of $2M$ timeslices. 

The ground state comparison shows general agreement between models for large timeslices, but as expected, the models with larger $M$ make a better variational estimate of the ground state at early times because the excited state contamination is absorbed into the excited states.  The first excited state comparison shows that the $M=2$ extraction is significantly different than the extractions obtained from $M=3,4,5$. Compared to the models with more states, the $M=2$ state extraction does not make a good variational estimate of the first excited state.

In order to test the validity of the extractions from Prony's method, we compare the effective masses to the masses obtained from a $\chi^2$ minimization fit to a sum of decaying exponentials. Here, the data is fit to a sum of three states or four states, for various different fitting regions. The $\chi^2$ values per degree of freedom for each of these fits is summarized in Table~\ref{chi_table} in the appendix. Figs.~\ref{box_ground_M2}-~\ref{box_second_M5} provide a qualitative comparison between the effective masses found via Prony's method, and the masses found through exponential fitting. In each plot, the colored bands represent the best fit values of exponential fits to the correlation function for the different fitting regions. The height of the band is given by $1\sigma$ jackknife errors. To visualize the joint distribution of masses over all of the timeslices used in the Prony approach, we overlay error boxes such that the darkest regions in the plot represent the joint estimate of the mass for the given model. Since a model using $M$ states requires a stencil of $2M$ timeslices, each point on the effective mass plots above has been extended to a box, where the vertical extent represents the error, and the horizontal extent of the box represents the stencil of timeslices used to compute that mass and error. For each timeslice, we overlay the (10,90), (20,80), (30,70), (40,60) percentile error boxes such that a region's darkness on these box plots represents the certainty that the true mass lies within that region.

Figs.~\ref{box_ground_M2}-~\ref{box_ground_M5} show the box plots for the $M=2,3,4,5$ state extractions of the ground state. Based on the darkest regions of the plots, the Prony estimates of the ground state show modest qualitative agreement with the estimates from the exponential fits. However, note from the scale of the y-axis, the precision of the ground state improves for larger values of $M$. For example, for $M=3$, the dark band ranges from about 0.077 to 0.084, while for $M=4$, the band ranges from about 0.081 to 0.085. Although the exponential fits are significantly more precise for the ground state, for the first and second excited states, the certainties are comparable between the two methods. The first excited state is clearly best resolved in the $M=4$ model, where the certainty band ranges from about 0.27 to 0.33. This result is comparable to the systematic uncertainty in the least squares fits to three states, as indicated by the variation of the best-fit value over the fitting regions.

\begin{figure*}
        \centering
        \begin{subfigure}[b]{0.49\textwidth}
            \centering
            \includegraphics[width=1\textwidth]{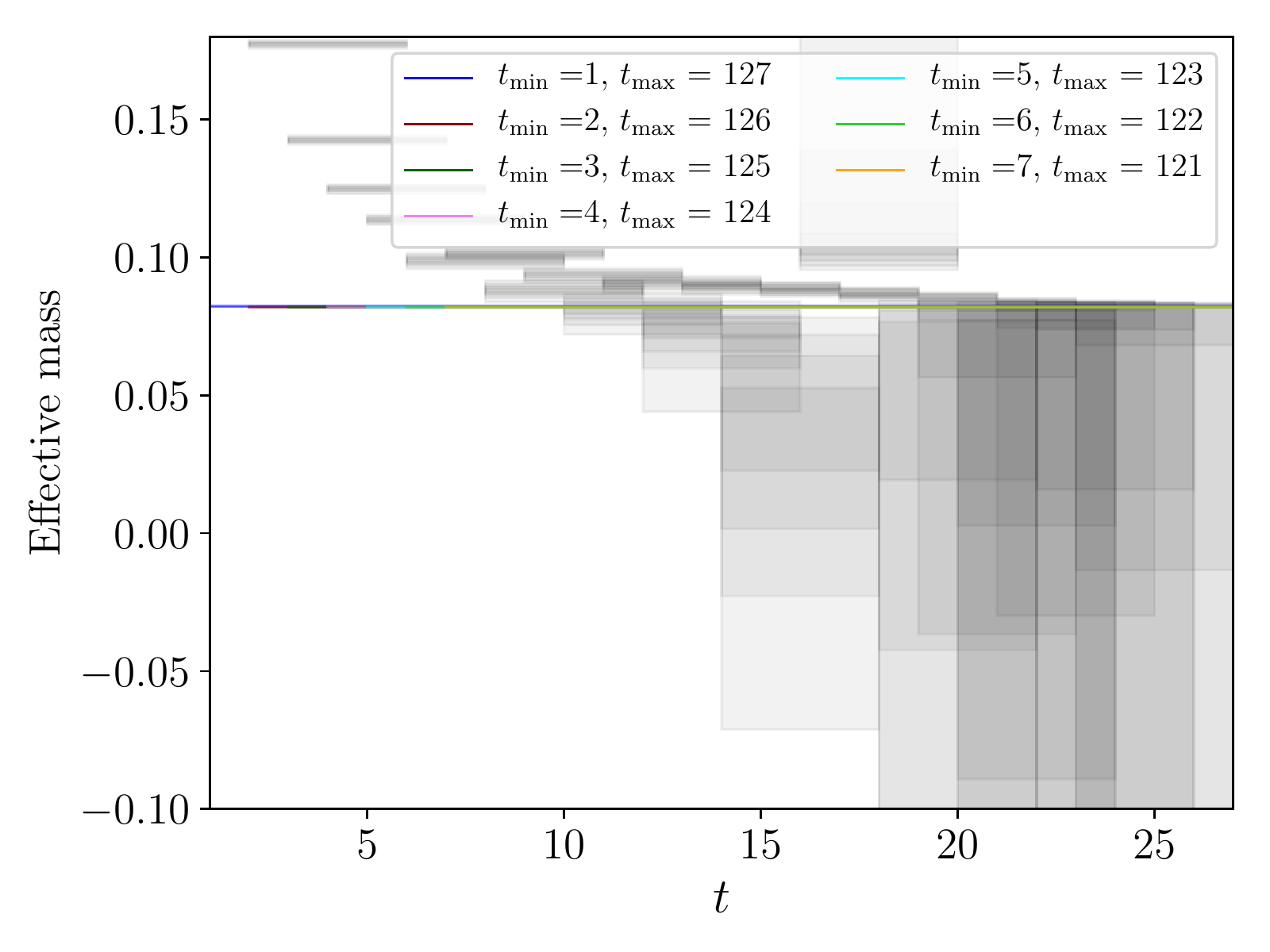}
            \caption{Exponential fit to three states.}    
        \end{subfigure}
        \hfill
        \begin{subfigure}[b]{0.49\textwidth}
            \centering
            \includegraphics[width=1\textwidth]{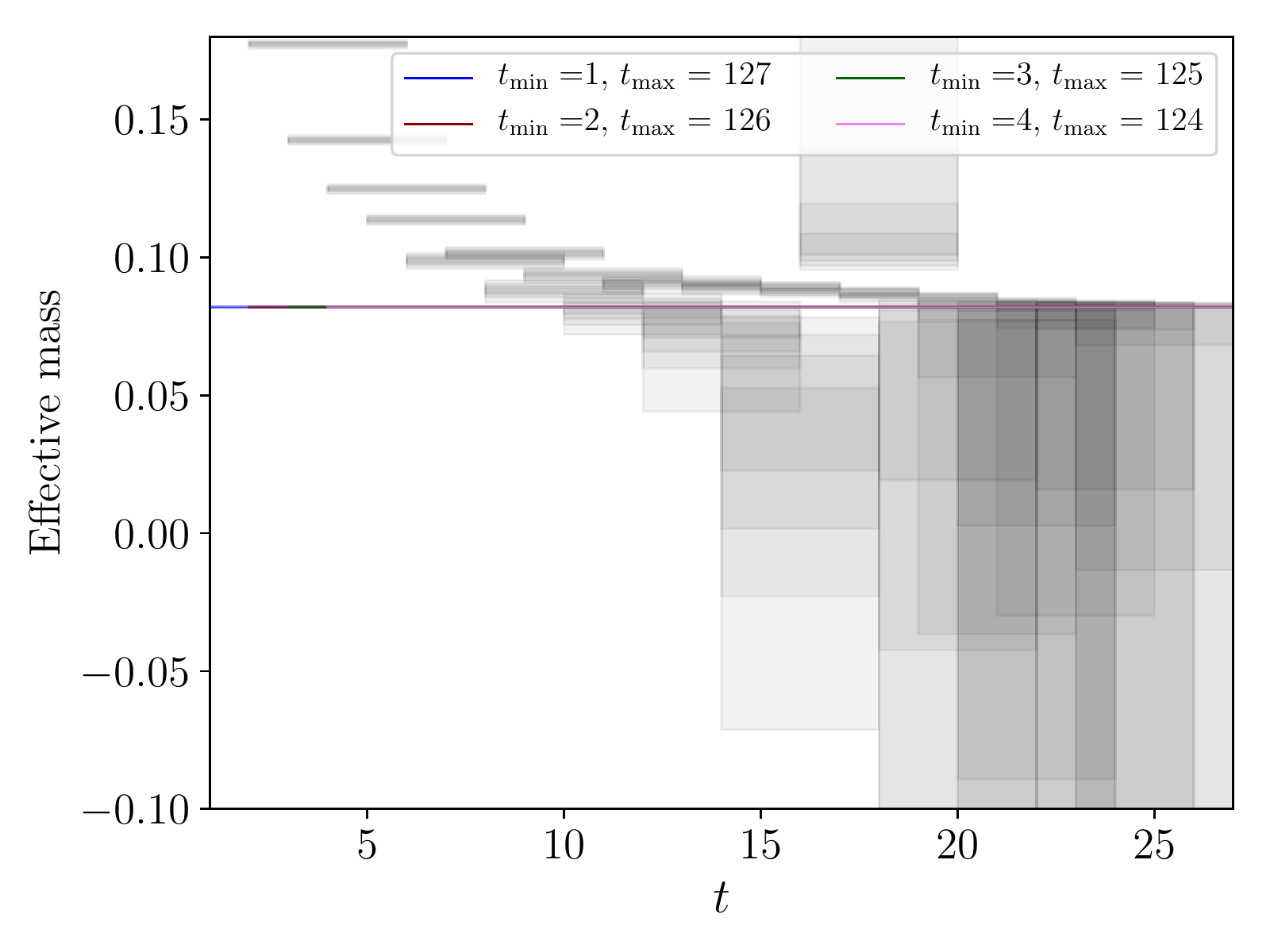}
            \caption{Exponential fit to four states.}    
        \end{subfigure}
	\caption{Ground state extraction from $M=2$ model.}
        \label{box_ground_M2}
\end{figure*}
\begin{figure*}
        \begin{subfigure}[b]{0.49\textwidth}  
            \centering 
            \includegraphics[width=1\textwidth]{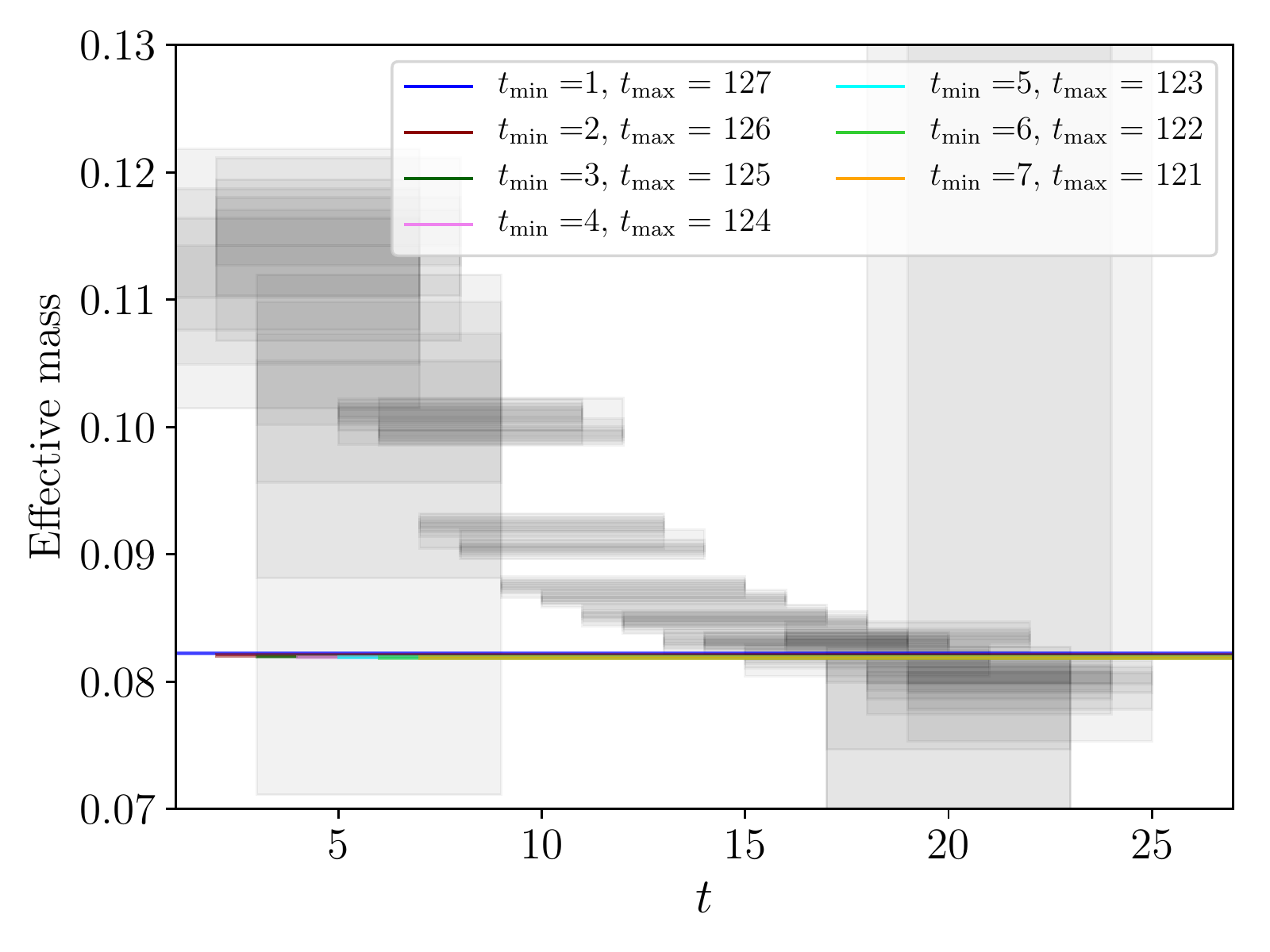}
            \caption{Exponential fit to three states.}  
        \end{subfigure}
        \hfill
        \begin{subfigure}[b]{0.49\textwidth}  
            \centering 
            \includegraphics[width=1\textwidth]{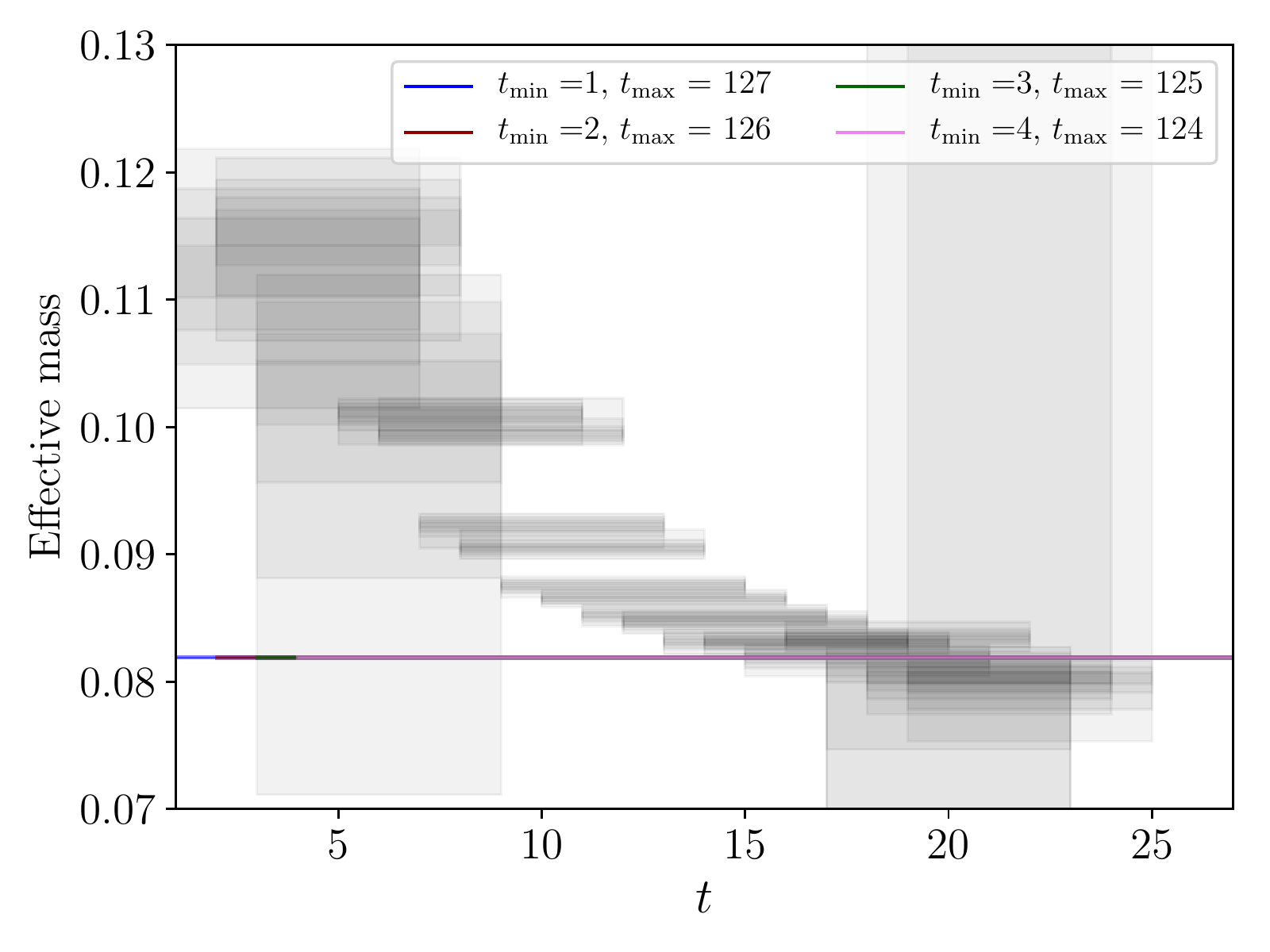}
            \caption{Exponential fit to four states.}  
        \end{subfigure}
        \caption{Ground state extraction from $M=3$ model.}
        \label{box_ground_M3}
\end{figure*}
\begin{figure*}
        \begin{subfigure}[b]{0.49\textwidth}   
            \centering 
            \includegraphics[width=1\textwidth]{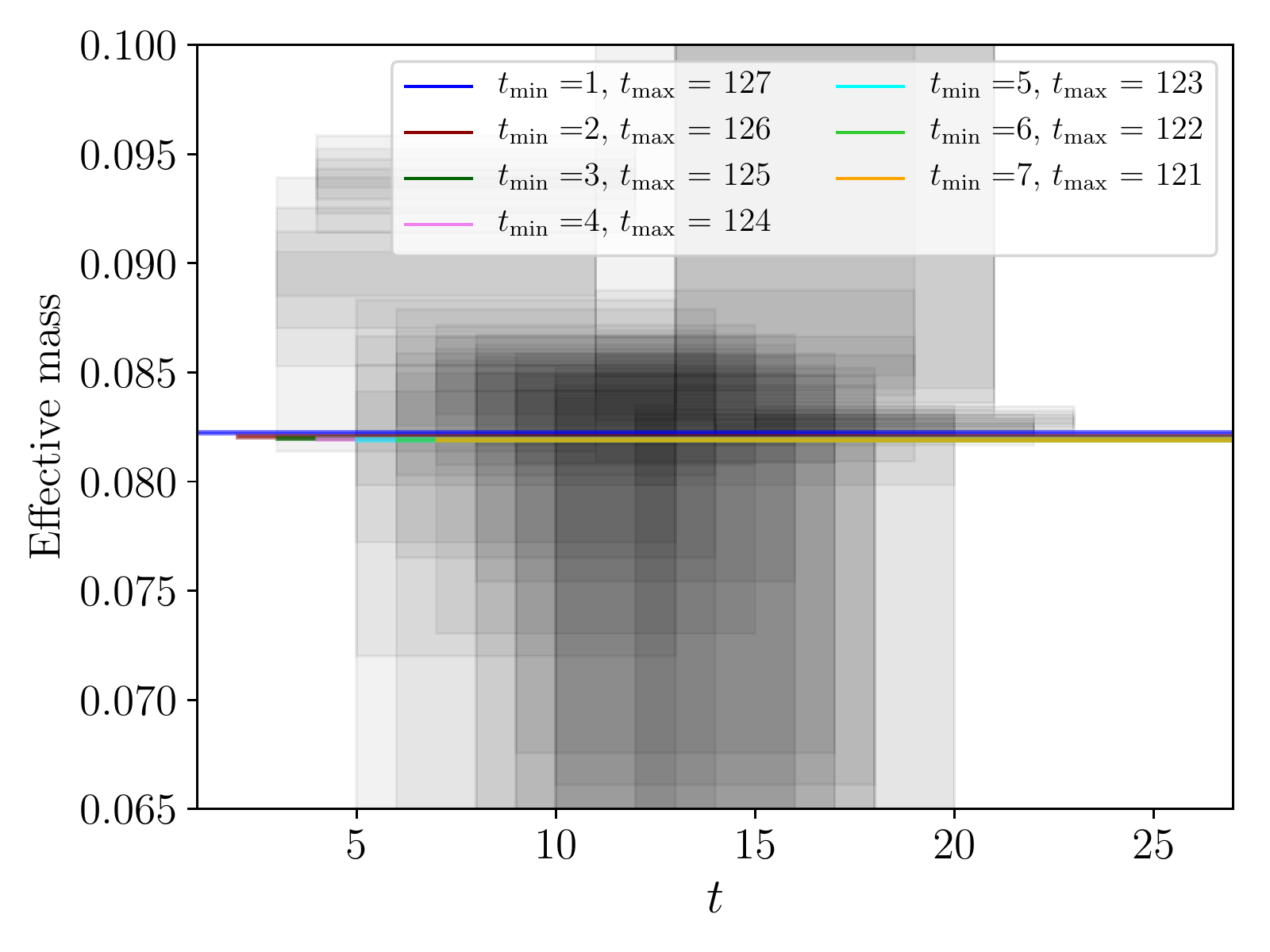}
            \caption{Exponential fit to three states.}    
        \end{subfigure}
        \hfill
        \begin{subfigure}[b]{0.49\textwidth}   
            \centering 
            \includegraphics[width=1\textwidth]{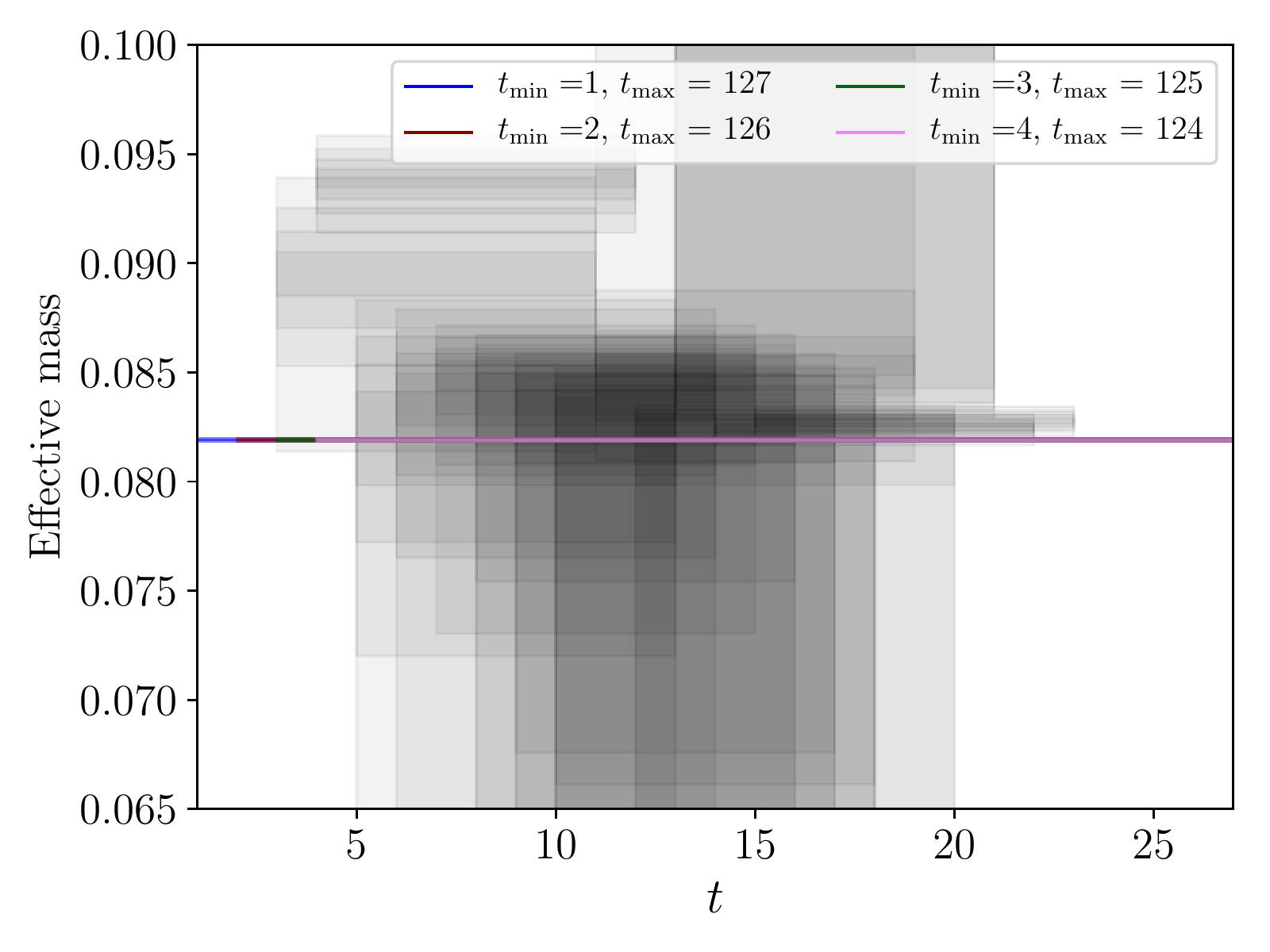}
            \caption{Exponential fit to four states.}    
        \end{subfigure}
          \caption{Ground state extraction from $M=4$ model.}
	 \label{box_ground_M4}
\end{figure*}
  \begin{figure*}
        \begin{subfigure}[b]{0.49\textwidth}   
            \centering 
            \includegraphics[width=1\textwidth]{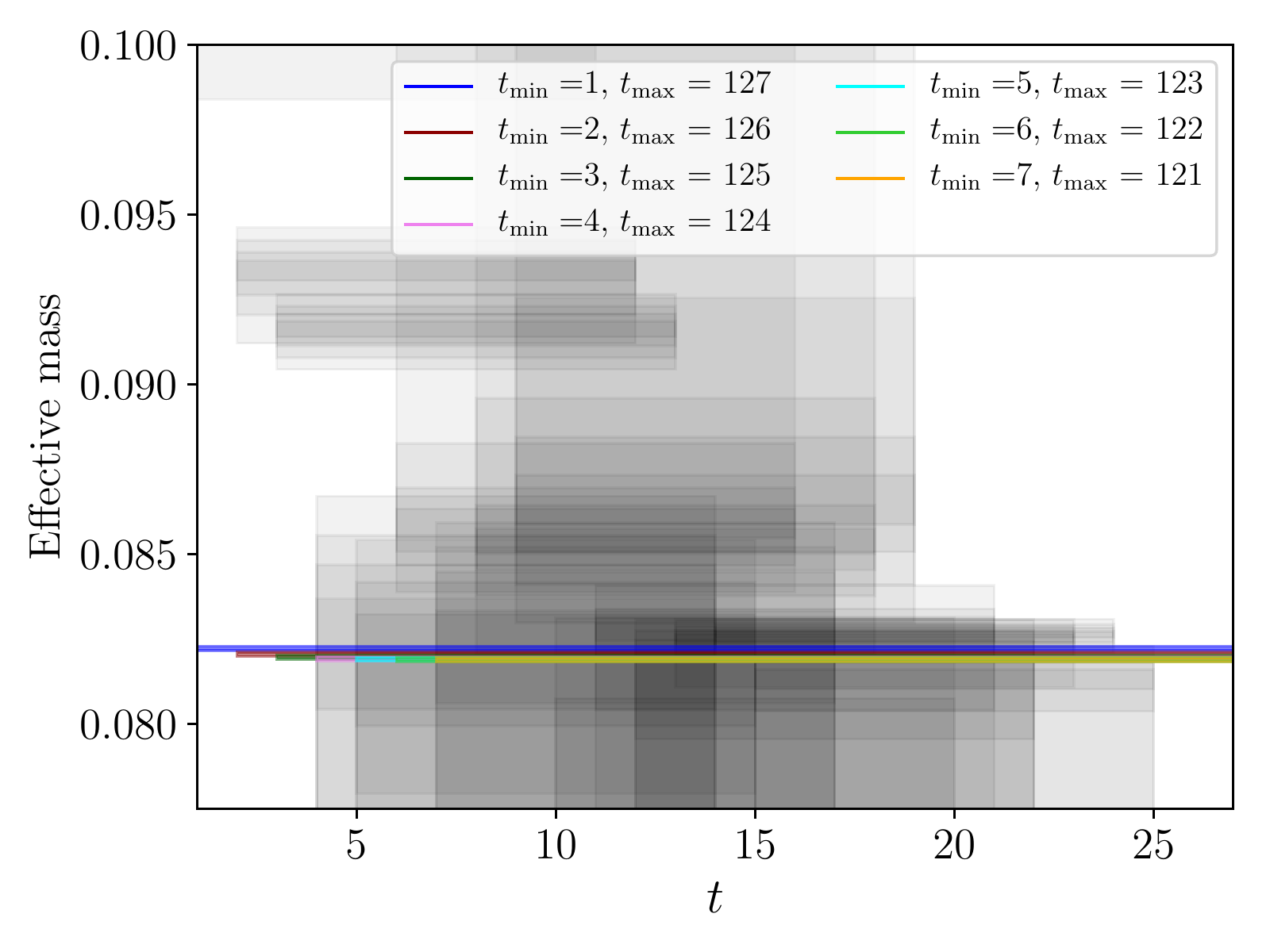}
            \caption{Exponential fit to three states.}    
        \end{subfigure}
        \hfill
        \begin{subfigure}[b]{0.49\textwidth}   
            \centering 
            \includegraphics[width=1\textwidth]{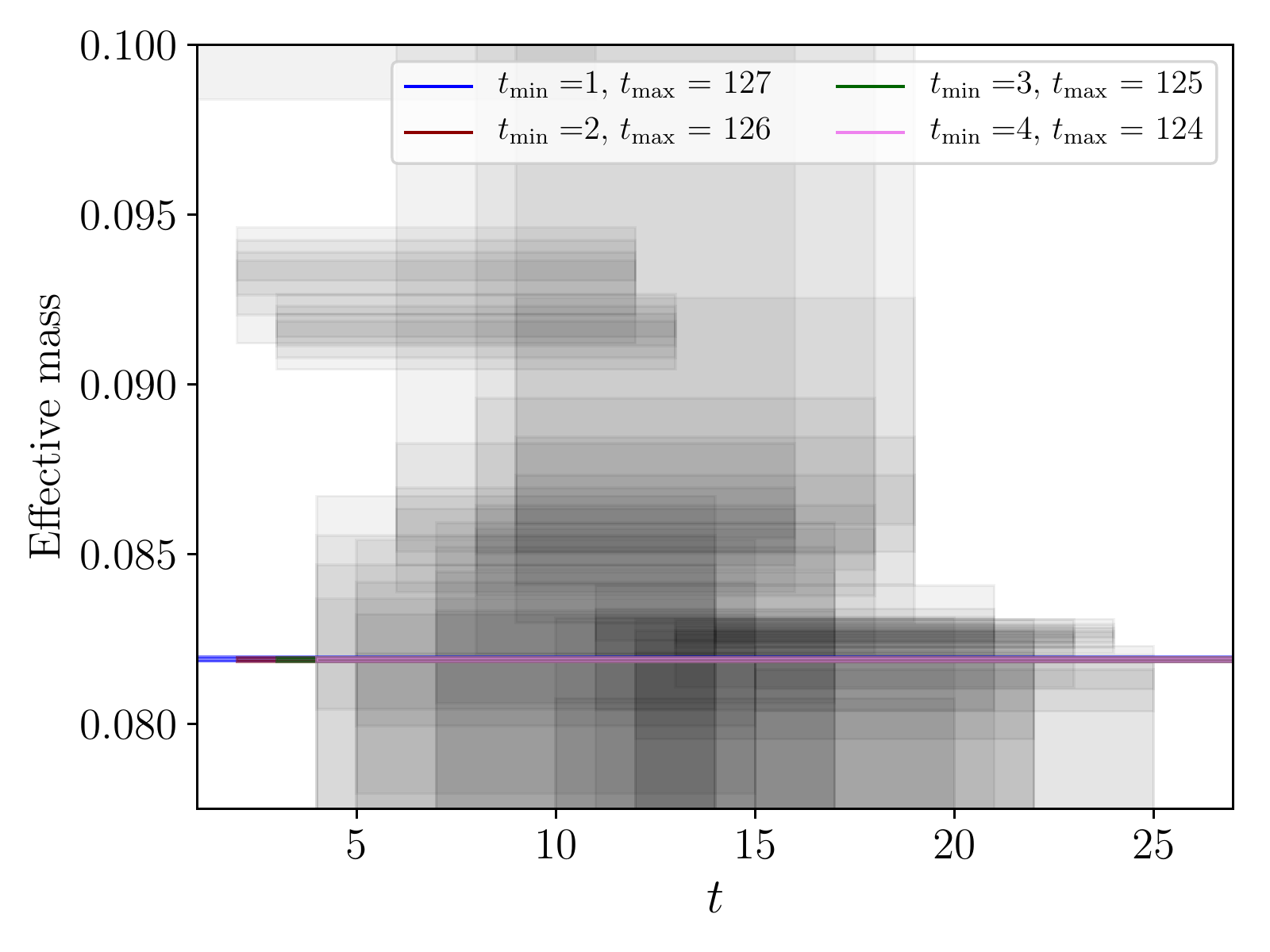}
            \caption{Exponential fit to four states.}    
        \end{subfigure}
  	\caption{Ground state extraction from $M=5$ model.}
	 \label{box_ground_M5}
\end{figure*}
    \begin{figure*}
        \begin{subfigure}[b]{0.49\textwidth}   
            \centering 
            \includegraphics[width=1\textwidth]{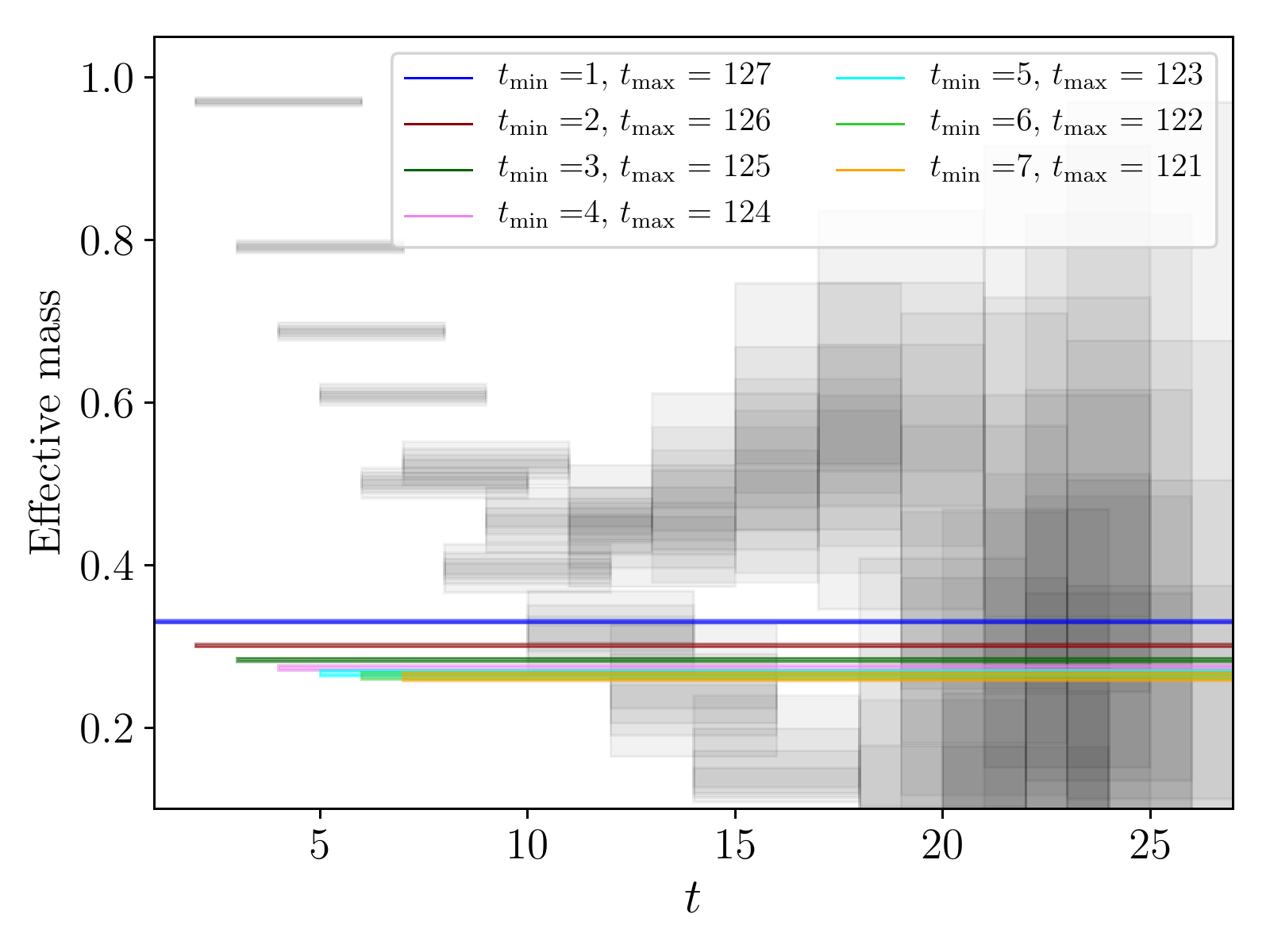}
            \caption{Exponential fit to three states.}    
        \end{subfigure}
        \hfill
         \begin{subfigure}[b]{0.49\textwidth}   
            \centering 
            \includegraphics[width=1\textwidth]{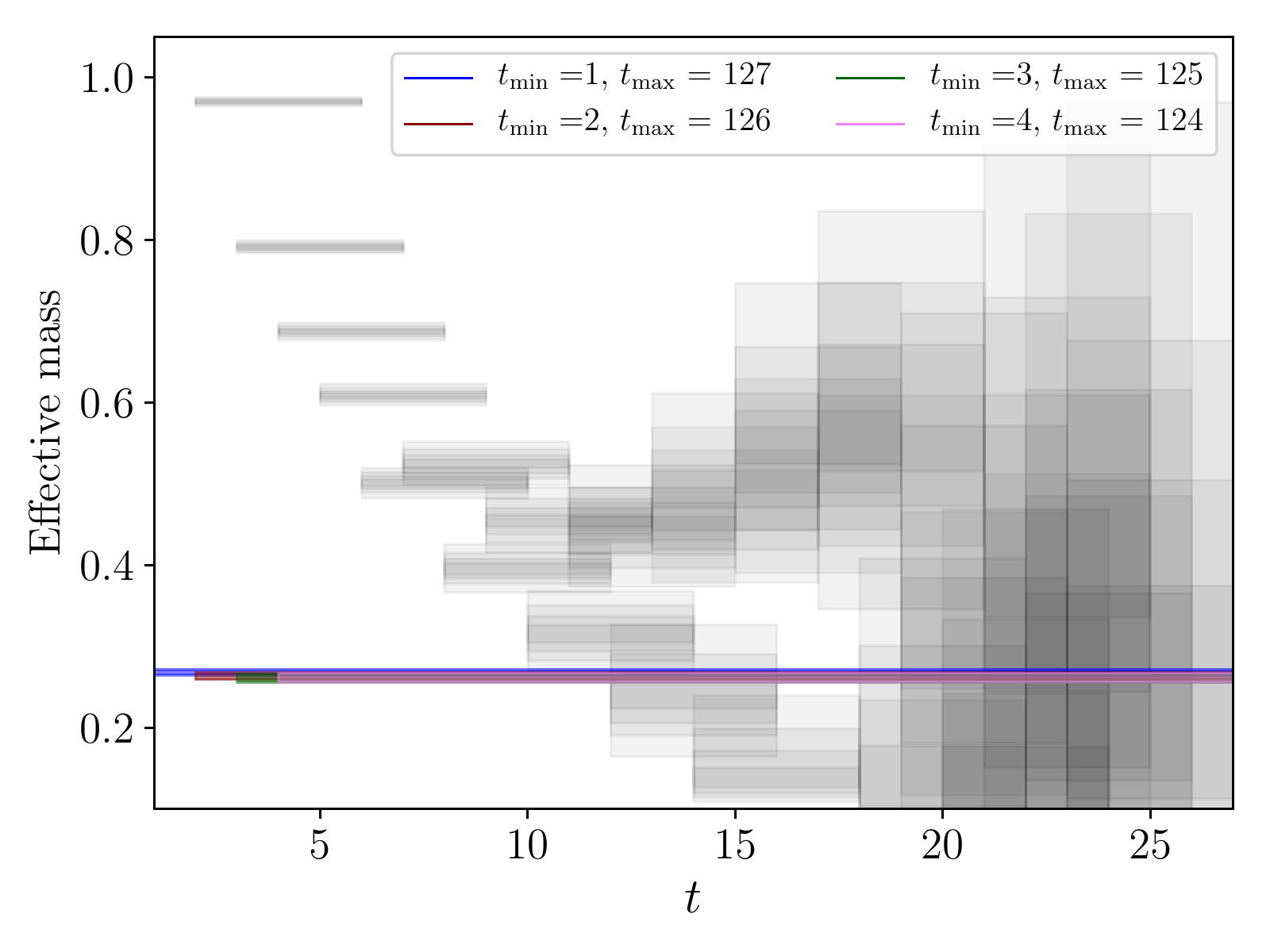}
            \caption{Exponential fit to four states.}    
        \end{subfigure}
	\caption{First excited state extraction from $M=2$ model.}
	 \label{box_first_M2}
\end{figure*}
    \begin{figure*}
        \begin{subfigure}[b]{0.49\textwidth}   
            \centering 
            \includegraphics[width=1\textwidth]{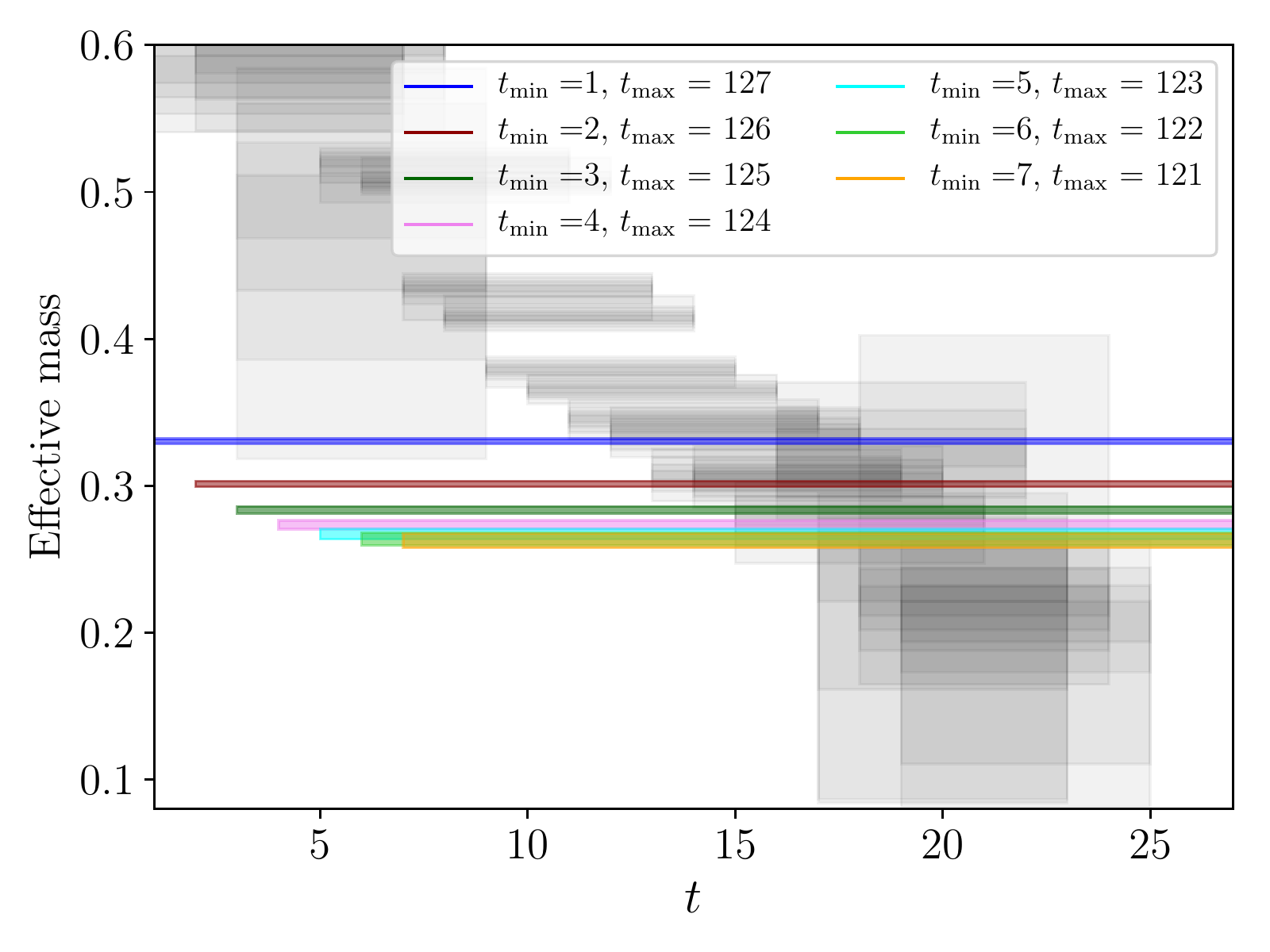}
            \caption{Exponential fit to three states.}    
        \end{subfigure}
        \hfill
         \begin{subfigure}[b]{0.49\textwidth}   
            \centering 
            \includegraphics[width=1\textwidth]{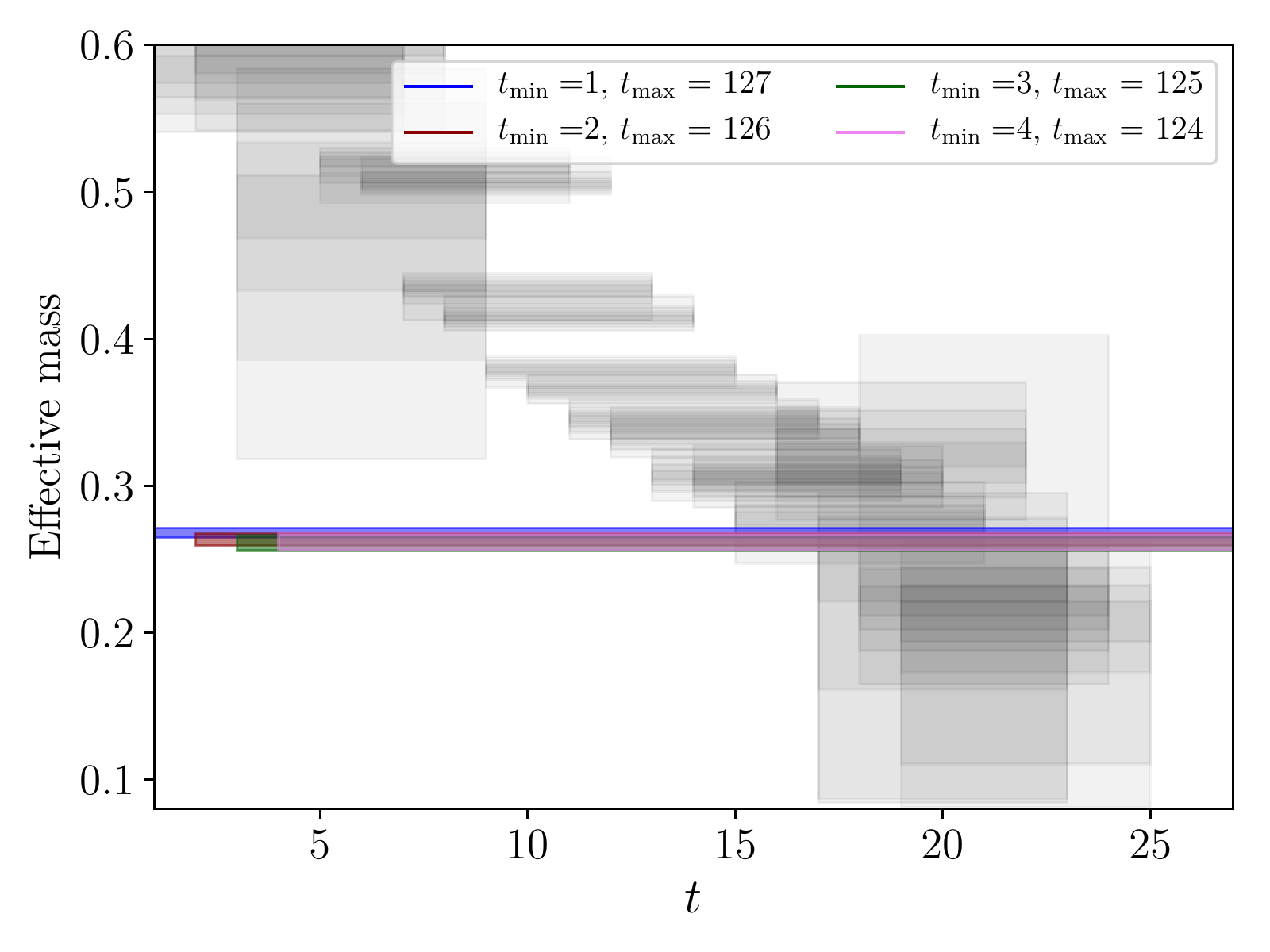}
            \caption{Exponential fit to four states.}    
        \end{subfigure}
	\caption{First excited state extraction from $M=3$ model.}
	 \label{box_first_M3}
\end{figure*}
    \begin{figure*}
        \begin{subfigure}[b]{0.49\textwidth}   
            \centering 
            \includegraphics[width=1\textwidth]{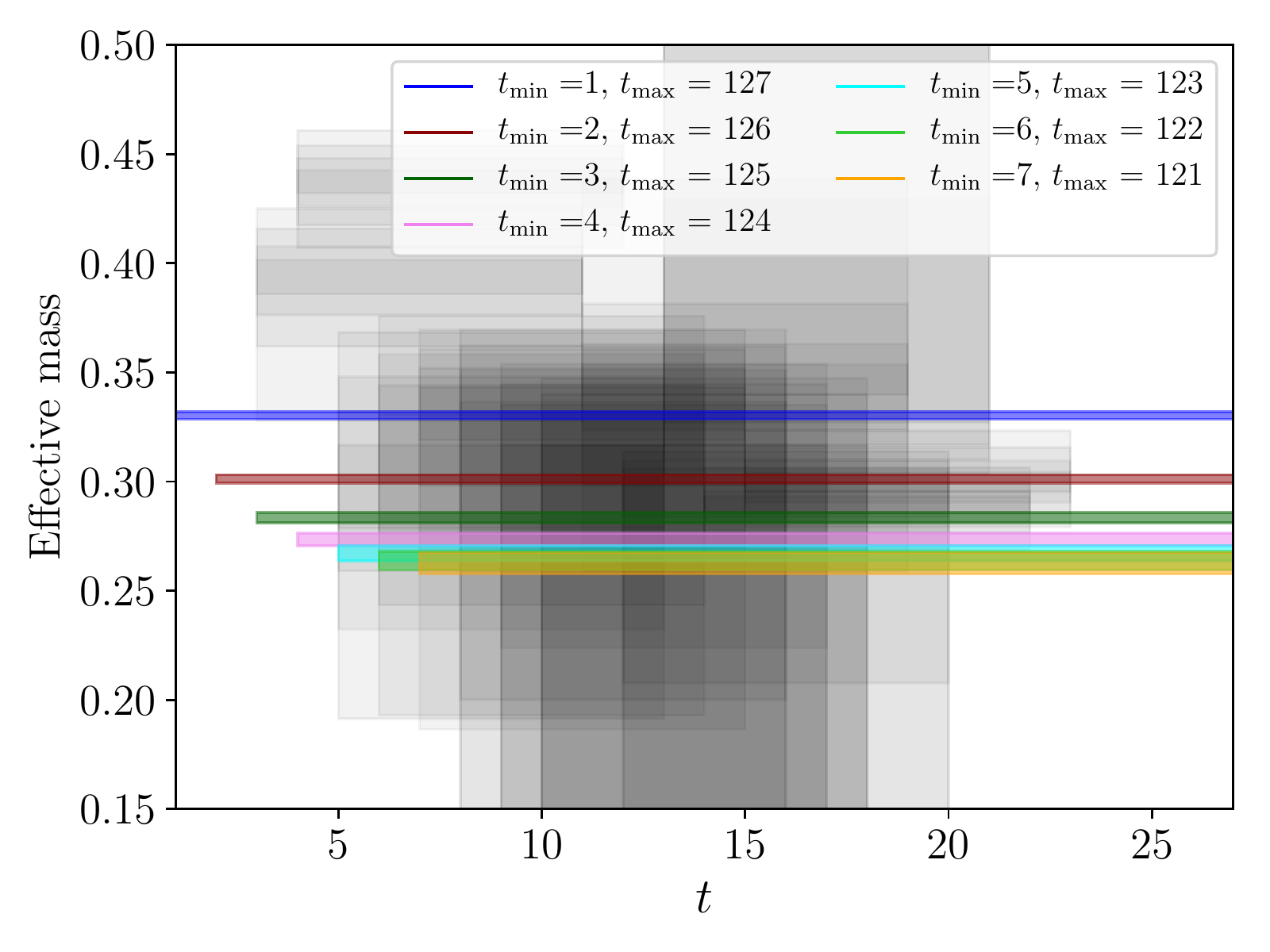}
             \caption{Exponential fit to three states.}    
        \end{subfigure}
        \hfill
         \begin{subfigure}[b]{0.49\textwidth}   
            \centering 
            \includegraphics[width=1\textwidth]{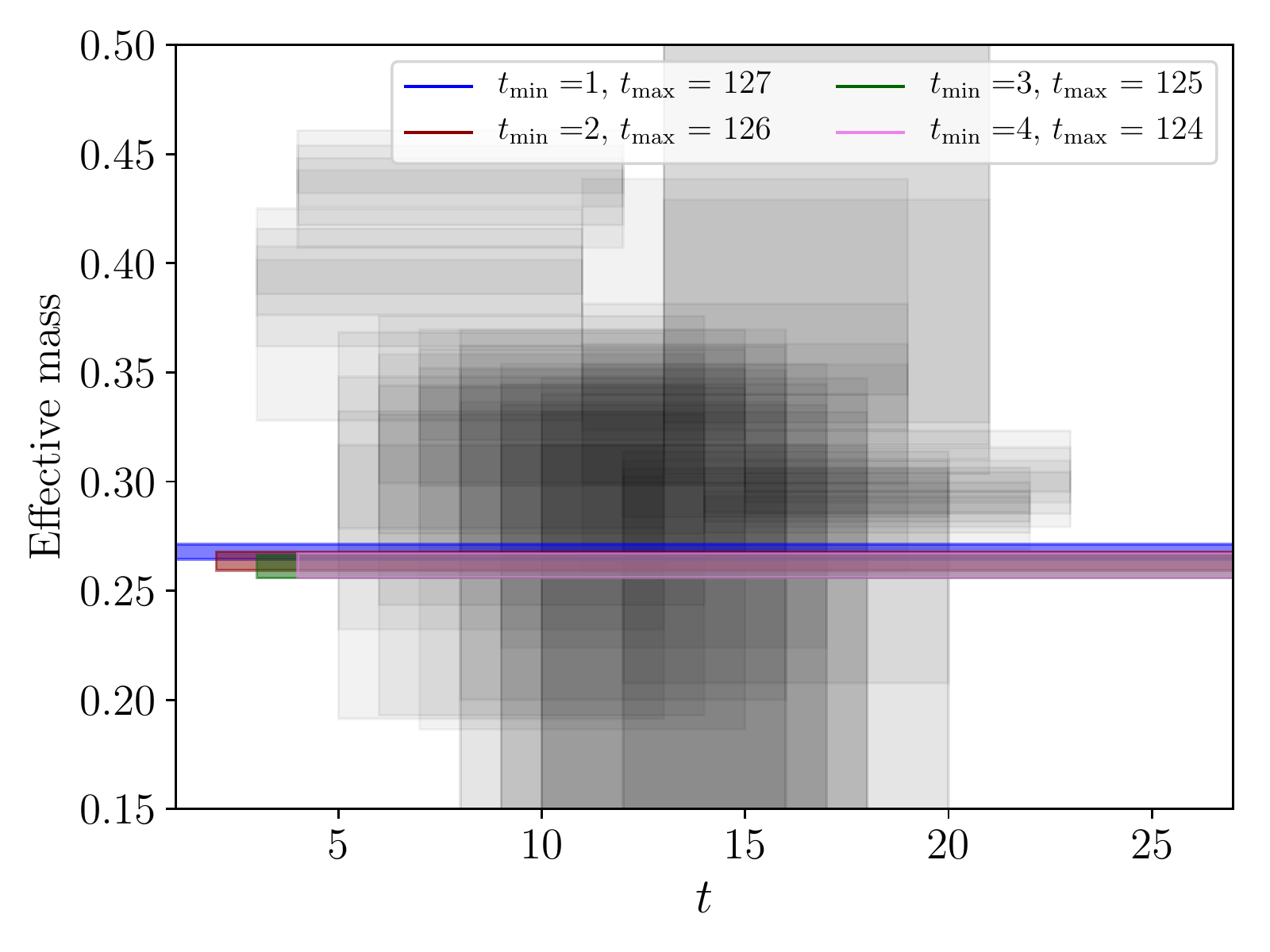}
             \caption{Exponential fit to four states.}    
        \end{subfigure}
	\caption{First excited state extraction from $M=4$ model.}
	 \label{box_first_M4}
\end{figure*}
    \begin{figure*}
        \begin{subfigure}[b]{0.49\textwidth}   
            \centering 
            \includegraphics[width=1\textwidth]{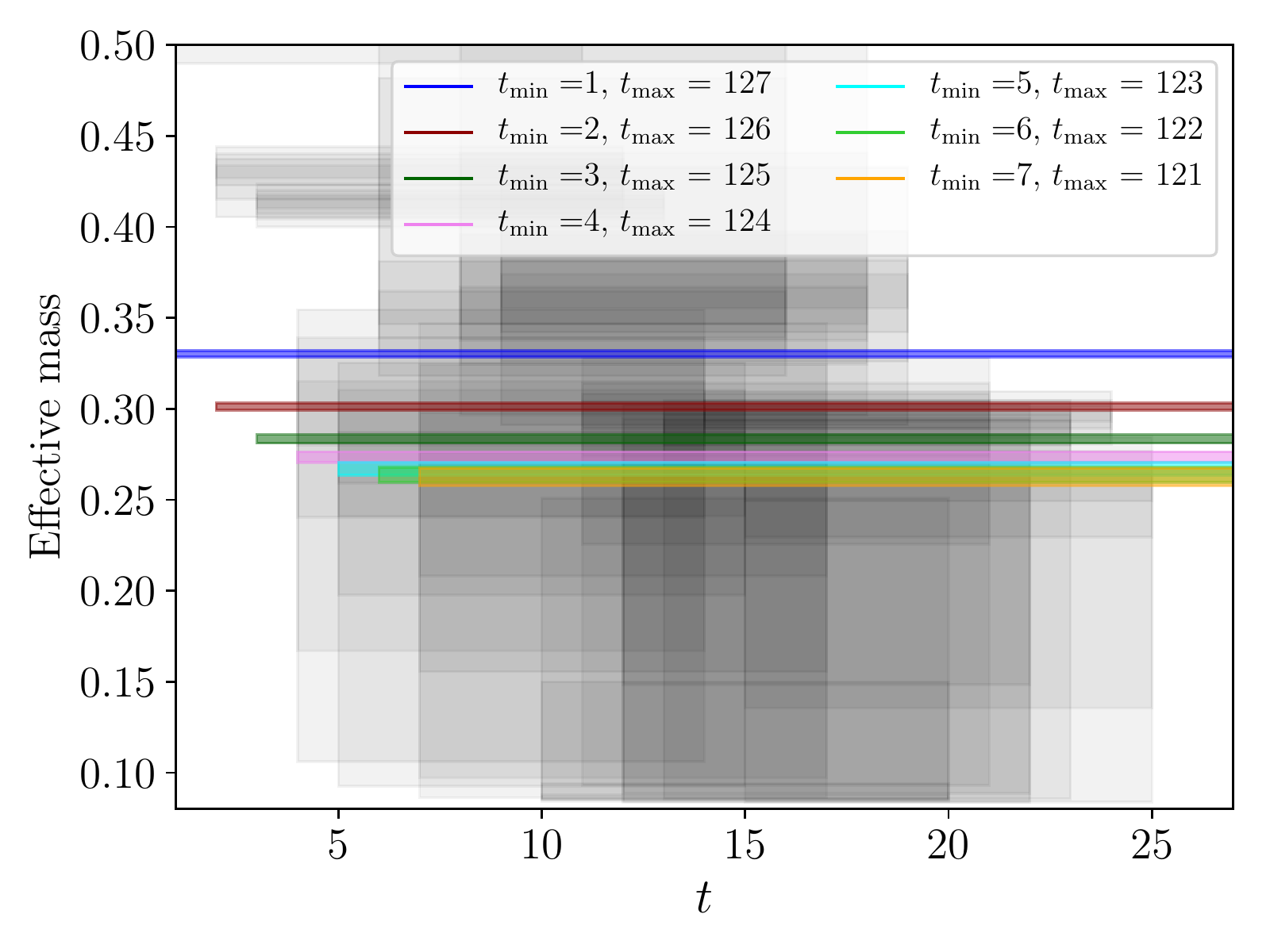}
            \caption{Exponential fit to three states.}    
        \end{subfigure}
        \hfill
         \begin{subfigure}[b]{0.49\textwidth}   
            \centering 
            \includegraphics[width=1\textwidth]{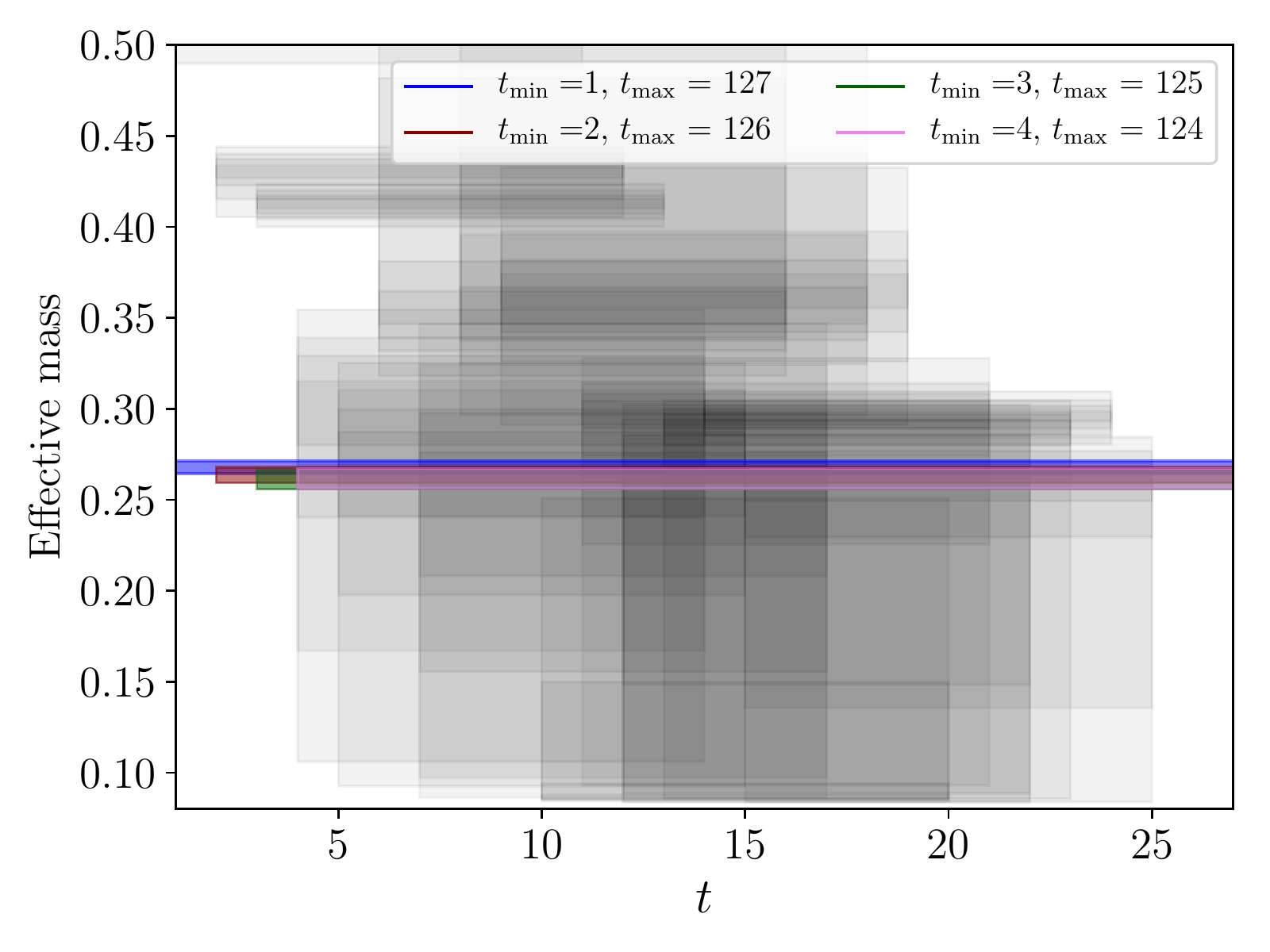}
            \caption{Exponential fit to four states.}    
        \end{subfigure}
        	\caption{First excited state extraction from $M=5$ model.}
         \label{box_first_M5}
 \end{figure*}
        \begin{figure*}
        \centering
        \begin{subfigure}[b]{0.49\textwidth}   
            \centering 
            \includegraphics[width=1\textwidth]{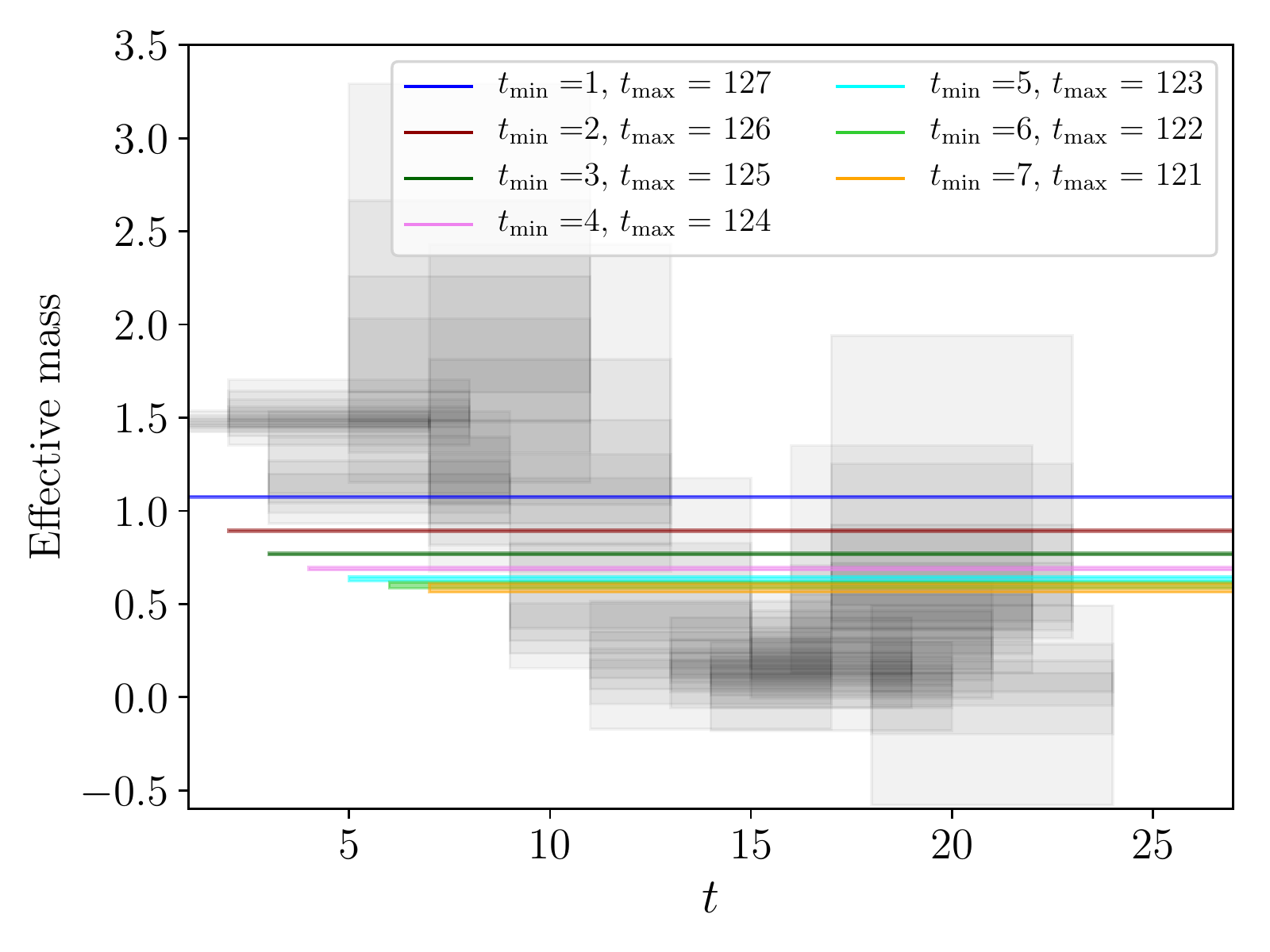}
            \caption{Exponential fit to three states.}    
        \end{subfigure}
        \hfill
        \begin{subfigure}[b]{0.49\textwidth}   
            \centering 
            \includegraphics[width=1\textwidth]{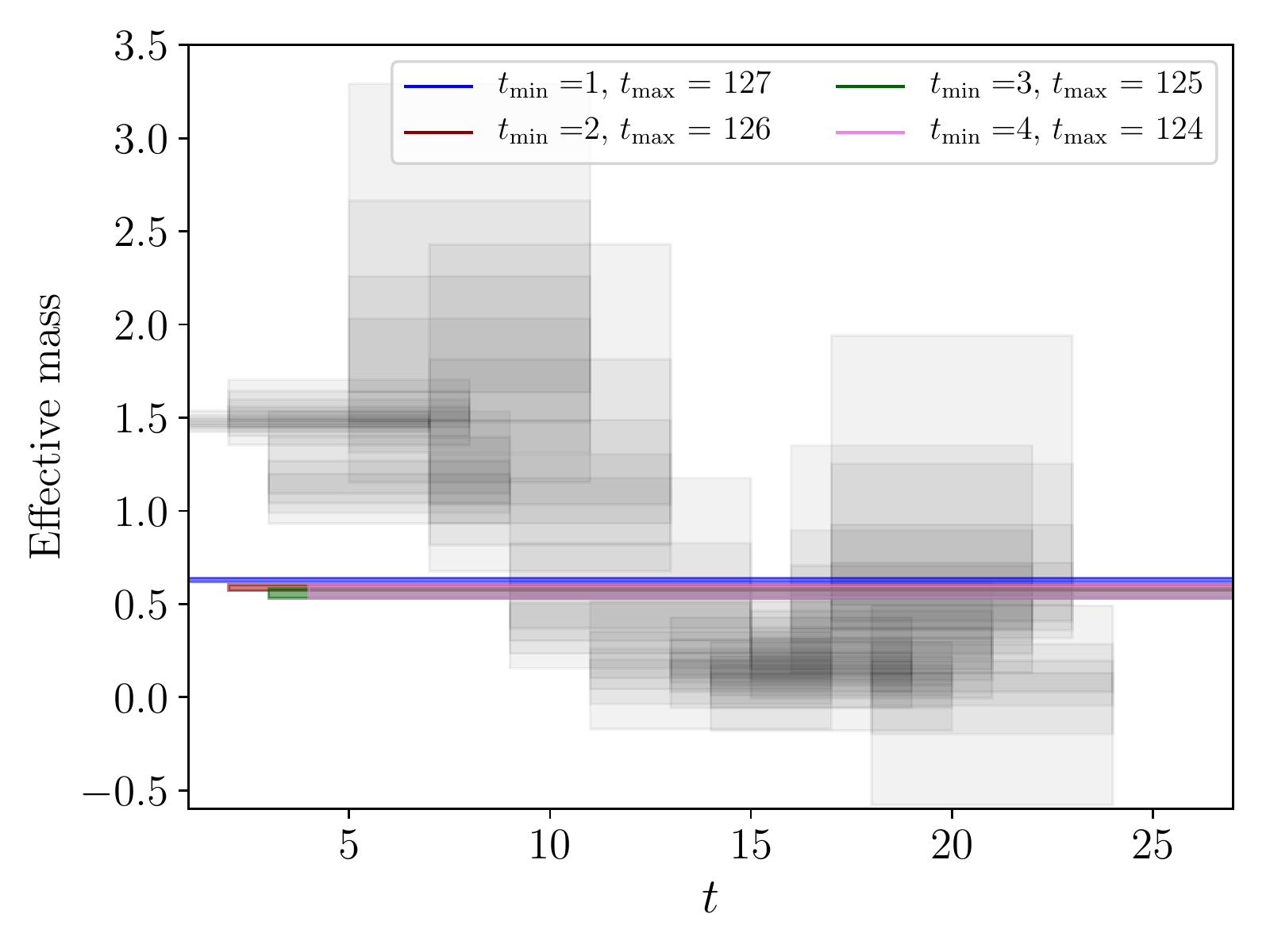}
            \caption{Exponential fit to four states.}    
        \end{subfigure}
        	\caption{Second excited state extraction from $M=3$ model.}
         \label{box_second_M3}
 \end{figure*}
        \begin{figure*}
        \begin{subfigure}[b]{0.49\textwidth}   
            \centering 
            \includegraphics[width=1\textwidth]{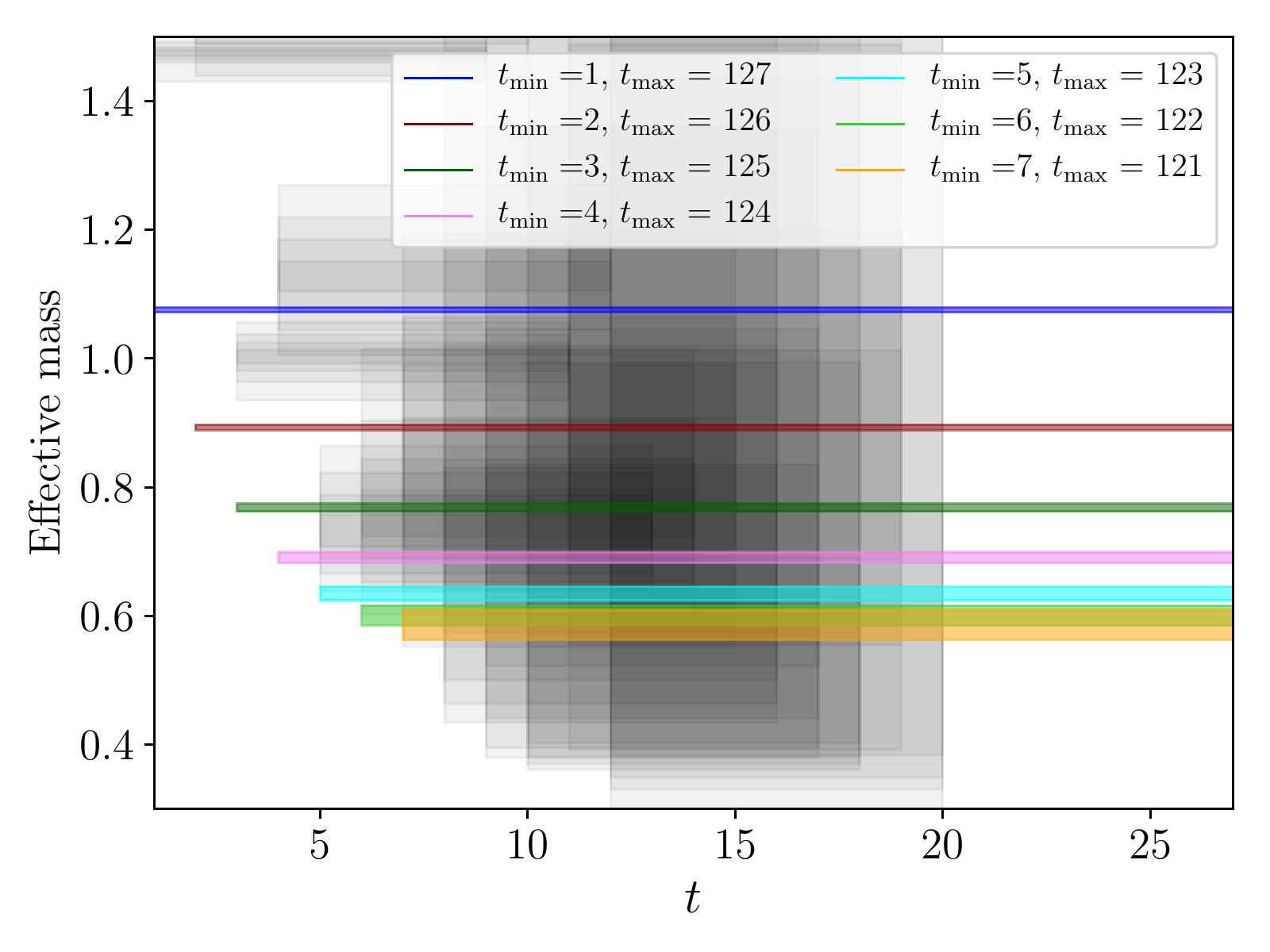}
            \caption{Exponential fit to three states.}    
        \end{subfigure}
        \hfill
        \begin{subfigure}[b]{0.49\textwidth}   
            \centering 
            \includegraphics[width=1\textwidth]{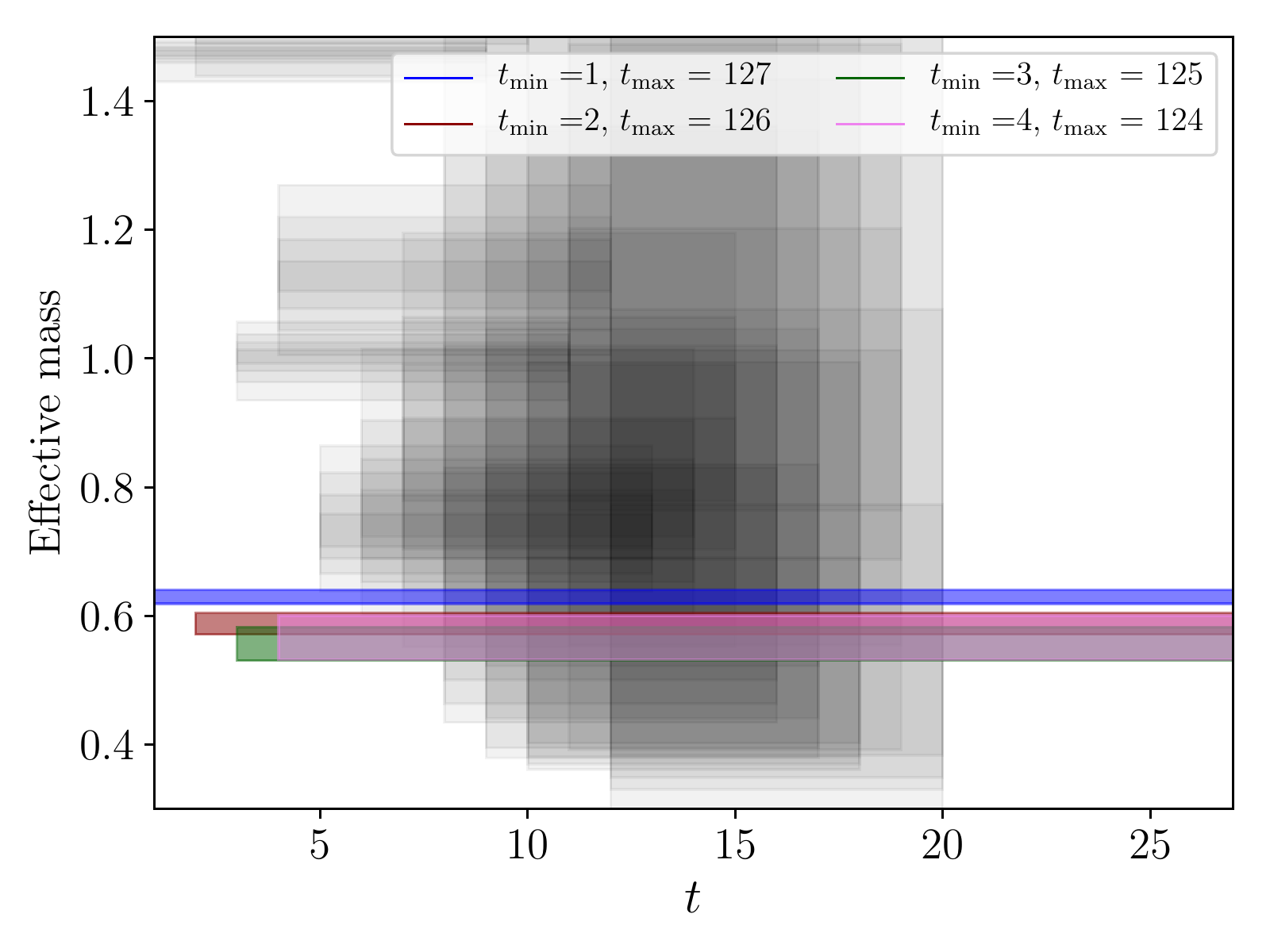}
            \caption{Exponential fit to four states.}    
        \end{subfigure}
       	\caption{Second excited state extraction from $M=4$ model.}
         \label{box_second_M4}
 \end{figure*}
        \begin{figure*}
        \begin{subfigure}[b]{0.49\textwidth}   
            \centering 
            \includegraphics[width=1\textwidth]{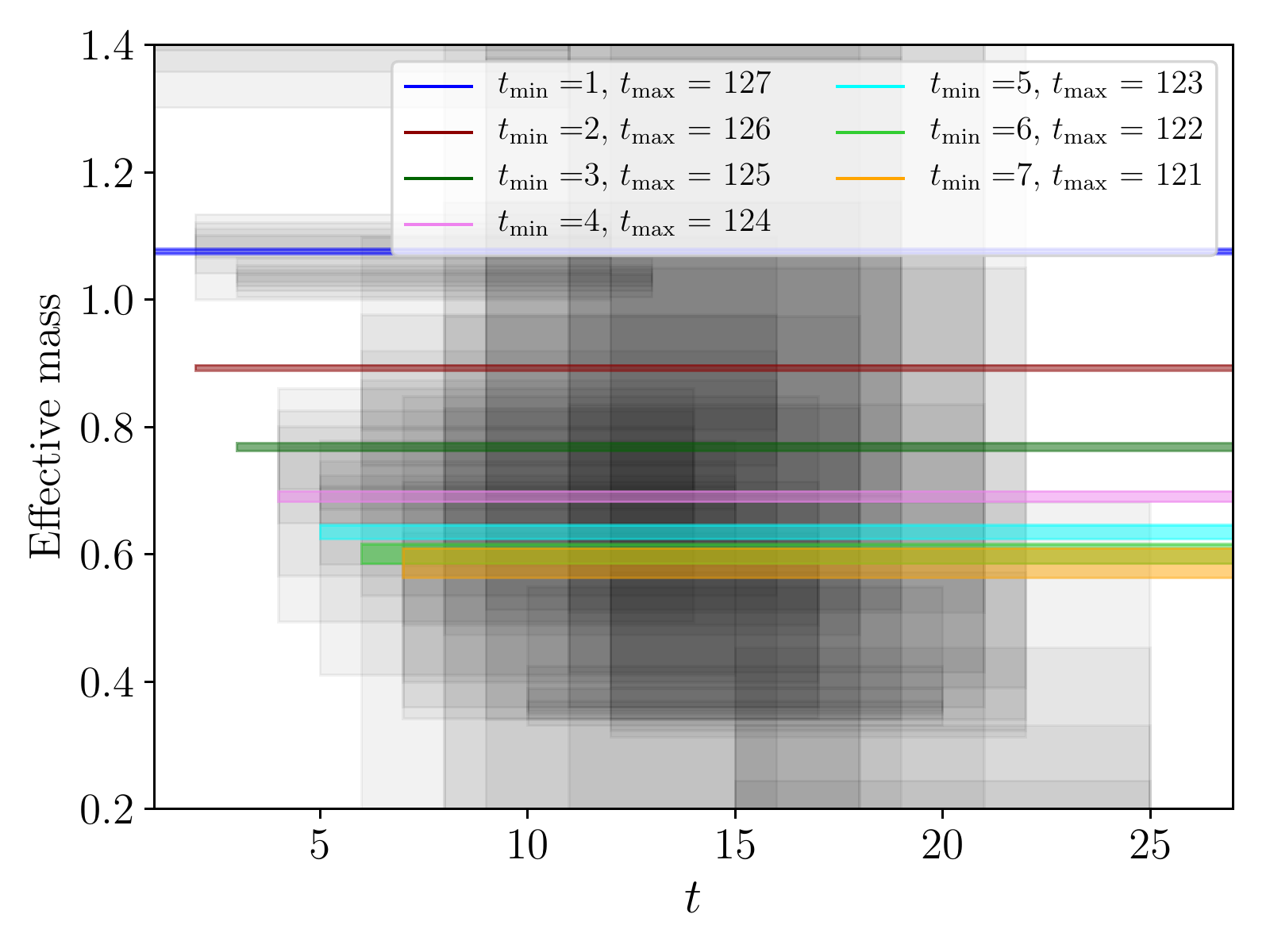}
            \caption{Exponential fit to three states.}    
        \end{subfigure}
        \hfill
         \begin{subfigure}[b]{0.49\textwidth}   
            \centering 
            \includegraphics[width=1\textwidth]{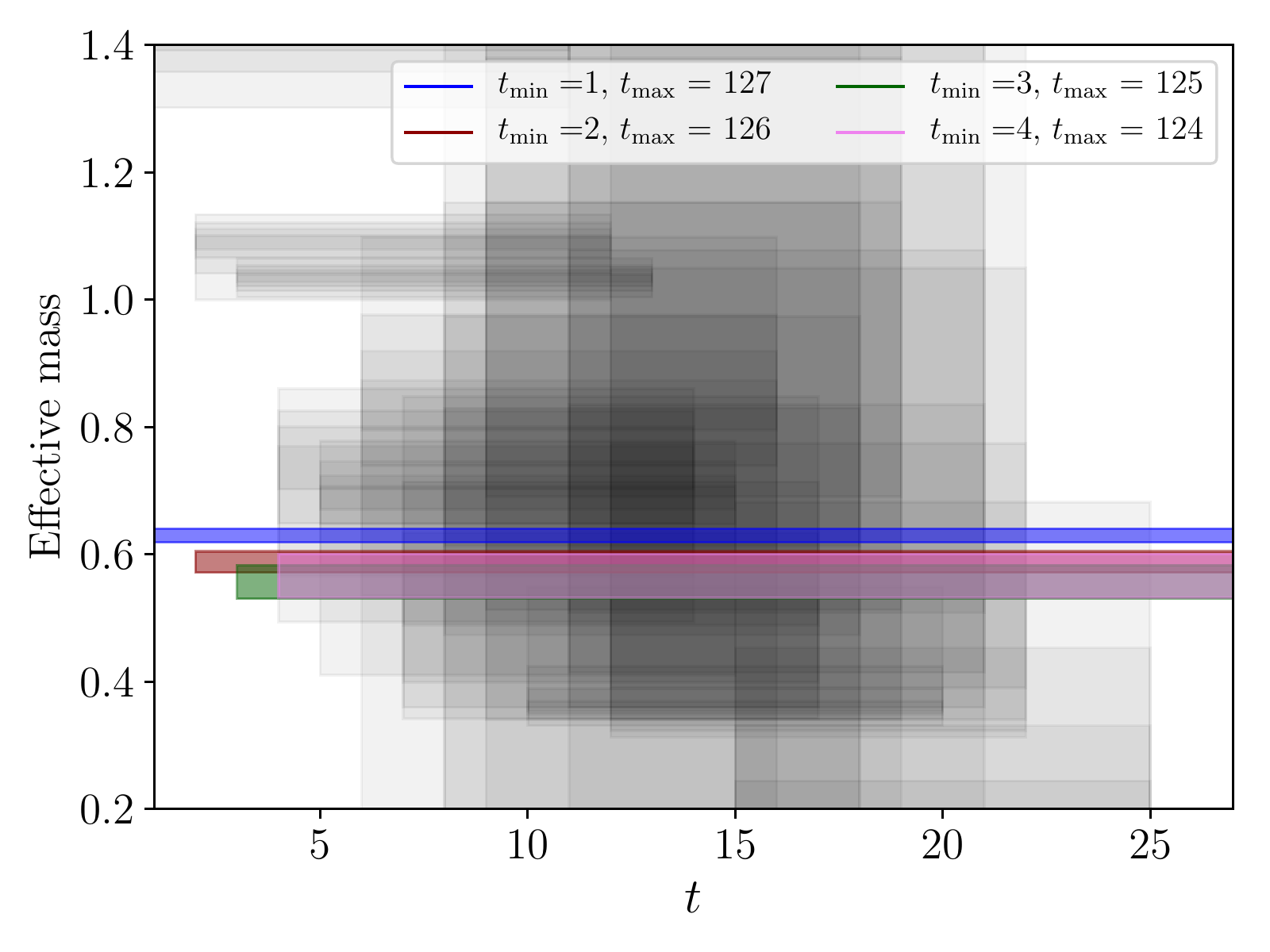}
            \caption{Exponential fit to four states.}    
        \end{subfigure}
    	\caption{Second excited state extraction from $M=5$ model.}
         \label{box_second_M5}
 \end{figure*}

In general, the three state fit contains significant systematic error for the different fitting regions, while the four state fit has much less systematic error due to the fitting regions because they agree within their statistical errors, which are larger for these fits. This indicates that the three state fit is insufficient to filter out excited state contamination. This can be compared to the time dependence of the Prony state extractions, indicated by where the uncertainty band lies along the time axis. For example, for the $M=4$ model extractions of the second excited state, it appears that the signal is strongest between timeslices $t=8$ to $t = 15$. Similar observations can be made by inspecting the $M=5$ model extraction of the second excited state. The mass of the second excited state from the three state fits shown here may have residual excited state contamination, and based on the systematically smaller best fit value of the mass for the different fitting regions.

\section{Conclusion}
In this work, we describe a new technique called automated label flows that allows for a systematic identification of states extracted from correlation functions in the presence of overlapping distributions. We illustrate this technique for Prony's method, but it can be easily adapted to any algebraic method for analyzing correlation functions, such as the variational method.

Our technique of automated label flows is reminiscent of techniques used to map out error regions in $\chi^2$ minimization fitting to sums of exponentials, where trajectories of curves are tracked within the $\chi^2$ error regions to obtain $\Delta \chi^2$ contours. However, in algebraic methods, where there is no analogous tunable $\chi^2$ value, we introduce a tunable parameter $\epsilon$ that characterizes the spread of the distribution of the data that is used in the algebraic method. Then, rather than flowing the fit parameters to different values of $\Delta \chi^2$, we flow the labels of bootstrap samples as $\epsilon$ increases.

In this work, we use single correlation functions in Prony's method, which results in the need to assign labels in a two dimensional parameter space of energies and amplitudes. However, Prony's method is easily generalizable to multiple correlation functions from the same quantum channel, which differ in their overlap with each state (see Ref.~\cite{Fleming:2009wb}). A set of $N$ correlation functions would yield a $N+1$ dimensional parameter space: one dimension for the energies, and $N$ dimensions for $N$ sets of amplitudes. One may naively think that in a higher dimensional space, the one dimensional trajectories of the automated label flows will be less likely to collide, or, similarly, the spectra resulting from another algebraic method such as the variational method would also be represented in a higher dimensional space, so label collisions are less likely to be a problem than what was observed here for Prony's method. However,  this is not necessarily the case because label collision occur in Prony's method whenever the discriminant of the polynomial equation which is solved to find the energies is less than or equal to zero. Hence, the mass extraction determines whether a label collision occurs, independent of the amplitudes, and under the influence of noise, the polynomial can exhibit complex conjugate pairs and therefore degeneracy. However, the behavior of the trajectories in higher dimensions is a worthwhile study of further investigations into automated label flows.

In the example data we show here, modest agreement is found between the extracted masses from Prony's method and those obtained from a least squares fit to sums of exponentials. Although the errors are much smaller for the ground state in the least squares method, our automated method provides a good estimation of excited states. Furthermore, Prony's method with automated label flows can be used to easily provide a least squares fitter with initial guesses that will be close to the best fit value, thus reducing the effort required to search the high dimensional space of a multi-exponential fit.

\begin{acknowledgments}
KC acknowledges support from the United States Department of Energy through the Computational Sciences Graduate Fellowship (DOE CSGF) through grant number DE-SC0019323. GF acknowledges support from the United States Department of Energy through grant number DE-SC0019061. We thank the LSD collaboration for allowing us to develop our method on a small subset of their correlation function data, and we hope to apply the analysis discussed in this work to more of their data in the future. 
\end{acknowledgments}

\appendix*
\section{}
In order to provide a fair comparison between the $\chi^2$ method and Prony's method with automated label flows, we provide the results of fits to the single pseudoscalar correlation function analyzed in this work. We note that this $\chi^2$ analysis is not as complete as the one described in the LSD publication~\cite{Appelquist:2018yqe} given that it only used one of the several available pion correlation functions and didn't fully account for data covariance. It is meant to only be used for a direct comparison with our Prony method which was done on the identical data set.

The values of $\chi^2$ per degree of freedom for the exponential fits shown if Figs.~\ref{box_ground_M2}-~\ref{box_second_M5} are given in Table~\ref{chi_table}, where the $\chi^2$ value is the jackknife average, and the error is jackknife error. Note that while the data is highly correlated, only the diagonal elements of the covariance matrix were used since there are only 232 correlators, so estimating the full covariance matrix is challenging. Hence, it is not surprising that the $\chi^2$ values are significantly less than $\chi^2=1$ per degree of freedom. However, the value of $\chi^2$ per degree of freedom is still a useful metric for the relative goodness of fit. From the point of view of $\chi^2$ minimization, each choice of fits in the table are comparable.  If $\chi^2$ was an absolute measure of goodness of fit, it could be used to help select the data range, $t_{\rm min}, t_{\rm max}$. Instead, a good fit can be determined by seeing that fit parameters are stable under varying $t_{\rm min}$ and $t_{\rm max}$, as depicted in Figs.~\ref{box_ground_M2}-~\ref{box_second_M5} for the four state fits. 

\begin{table*}
\centering
\begin{tabular}{|c|c|c|c|c|c|c|}
\hline
States in fit & $t_{\rm min},\, t_{\rm max}$&  ground state & first excited state & second excited state & third excited state &$\chi^2$/dof \\
\hline
3 & 1, 127   & 0.082227 (63) & 0.3303 (17)   & 1.0756 (38)  & n/a           &  0.51 (9) \\
\hline
3 & 2, 126   & 0.082056 (63) & 0.3012 (19)   &  0.8928 (44) & n/a           & 0.13 (7)\\
\hline
3 & 3, 125   & 0.081969 (63) & 0.2834 (24)   &  0.7686 (60) & n/a           & 0.05 (7)\\
\hline
3 & 4, 124   & 0.081930 (63) & 0.2735 (30)  &  0.6906 (82) & n/a            & 0.03 (7) \\
\hline
3 & 5, 124   & 0.081909 (63) & 0.2671 (35)  & 0.635 (11)    & n/a            & 0.03 (7)\\
\hline
3 & 6, 124   & 0.081900 (63) & 0.2637 (42)  & 0.601 (16)    & n/a            & 0.03 (7)\\
\hline
3 & 7, 124   & 0.081898 (64) & 0.2626 (49)  & 0.586 (23)    & n/a            & 0.03 (7)\\
\hline
4 & 1, 127   & 0.081912 (63) & 0.2679 (35)  & 0.630 (11)    & 1.441 (14) & 0.03 (7) \\
\hline
4 & 2, 126  & 0.081901 (64)  & 0.2636 (43)  & 0.588 (17)    & 1.291 (30) & 0.03 (7)\\
\hline
4 & 3, 125  & 0.081895 (64)  & 0.2611 (52)  & 0.557 (26)    & 1.147 (55) & 0.03 (7)\\
\hline
4 & 4, 124  & 0.081896 (64)  & 0.2616 (55)  & 0.566 (34)    & 1.22 (15)   & 0.03 (7)  \\
\hline
\end{tabular}
\caption{Fit values and $\chi^2$ per degree of freedom (dof) for each of exponential fits in Figs.~\ref{box_ground_M2}-~\ref{box_second_M5}.}
\label{chi_table}
\end{table*}

\bibliographystyle{unsrt}
\bibliography{Flows_v7}

\end{document}